%% file: main.tex
\documentclass[a4paper,11pt]{article}
\usepackage{a4,listings,fancyhdr}
\usepackage[dvips]{graphicx}
\usepackage{alltt,algorithmic,algorithm}
\usepackage{latexsym,amsmath,amssymb}
\usepackage{amsfonts}
\usepackage[latin1]{inputenc}
\usepackage[T1,OT1]{fontenc}
\usepackage{natbib}
\usepackage{setspace}   
\usepackage[gen]{eurosym}
\usepackage{threeparttable}
\usepackage[referable]{threeparttablex}
\usepackage{color}
\usepackage{subfigure}
\usepackage{rotating}
\usepackage{environ}
\usepackage{booktabs}
\usepackage{pdfpages}
\usepackage{pifont}
\usepackage[absolute]{textpos}
\usepackage{lineno}
\usepackage[hang,small,bf,singlelinecheck=false]{caption}
\usepackage{pdflscape}
\usepackage{comment}
\usepackage{mdframed}
\usepackage{soul}

\usepackage{longtable}
\usepackage[colorlinks,pdfusetitle]{hyperref}
\definecolor{darkred}{rgb}{0.9,0,0}
\definecolor{darkgreen}{rgb}{0,0.6,0}
\definecolor{darkblue}{rgb}{0,0,0.9}
\definecolor{lightblue}{RGB}{0,191,255}

\definecolor{darkgray}{RGB}{99, 99, 99}
\definecolor{darkgray2}{RGB}{194, 194, 194}
\definecolor{darkgray3}{RGB}{207, 207, 207}
\definecolor{darkgray4}{RGB}{232, 232, 232}
\definecolor{whitegrey}{RGB}{247, 247, 247}

\hypersetup{colorlinks,linkcolor=darkblue,filecolor=darkred,urlcolor=darkblue, citecolor=darkblue}

\usepackage[blocks]{authblk}
\usepackage[left=2cm,right=2cm,top=2cm,bottom=2cm,includeheadfoot]{geometry} %
\usepackage[flushmargin]{footmisc} %
\usepackage{setspace} %
\usepackage{sectsty}
\sectionfont{\normalsize}
\subsectionfont{\normalsize}
\subsubsectionfont{\normalsize}
\usepackage[small,compact]{titlesec}
\titlespacing\section{0pt}{6pt plus 2pt minus 2pt}{0pt plus 1pt minus 0pt}
\titlespacing\subsection{0pt}{6pt plus 2pt minus 2pt}{0pt plus 1pt minus 0pt}
\titlespacing\subsubsection{0pt}{6pt plus 2pt minus 2pt}{0pt plus 1pt minus 0pt}

\usepackage{listings}
\definecolor{codegreen}{rgb}{0,0.6,0}
\definecolor{codegray}{rgb}{0.5,0.5,0.5}
\definecolor{codepurple}{rgb}{0.58,0,0.82}
\definecolor{codeblack}{rgb}{0,0,0}
\definecolor{backcolour}{rgb}{1,1,1}

\lstdefinestyle{mystyle}{
    backgroundcolor=\color{backcolour},   
    commentstyle=\color{codegray},
    keywordstyle=\bfseries\color{codeblack},
    numberstyle=\tiny\color{codegray},
    stringstyle=\color{codeblack},
    basicstyle=\ttfamily\footnotesize,
    breakatwhitespace=false,         
    breaklines=true,                 
    captionpos=b,                    
    keepspaces=true,                 
    numbers=left,                    
    numbersep=5pt,                  
    showspaces=false,                
    showstringspaces=false,
    showtabs=false,                  
    tabsize=2
}

    \makeatletter
\def\@fnsymbol#1{\ensuremath{\ifcase#1\or *\or **\or
   \mathsection\or \mathparagraph\or \|\or **\or \dagger\dagger
   \or \ddagger\ddagger \else\@ctrerr\fi}}
    \makeatother

\begin{document}

\singlespacing

\title{AI and the Dynamic Supply of Training Data\thanks{Email: christian.peukert@unil.ch (corresponding author), florian.abeillon@proton.me, jeremie.haese@unil.ch, franziska.kaiser@business.uzh.ch,  alexander.staub@unil.ch. The first author led the project and contributed the most, while other authors who made significant contributions are listed alphabetically. We are grateful to Anthi Kiouka for her valuable input and discussions. We also thank Timothy Carbone at Unsplash for generously answering our questions, giving helpful insights and providing access to data. We thank Charles Ayoubi, Heski Bar-Isaac, Stefan Bechtold, Grazia Cecere, Raffaele Conti, Chiara Farronato, Joshua Gans, Avi Goldfarb, Christophe G\"osken, Shane Greenstein, Brad Greenwood, Ginger Zhe Jin, Reinhold Kesler, Mario Leccese, Sijie Lin, Megan MacGarvie, Ananya Sen, Tim Simcoe, Scott Stern, Daniel Sokol, Holger Spamann, Catherine Tucker, Pinar Yildirim, and Joel Waldfogel for helpful comments. The paper has benefited from the feedback of participants at seminars and conferences at Amsterdam Business School, ESSEC Business School, ETH Zurich (Center for Law and Economics), Google Economics, Harvard Business School (Digital Competition and Tech Regulation Conference and Open and User Innovation Conference), HEC Lausanne, NBER Summer Institute on Digital Economics and AI, NBER Productivity Seminar, Nova SBE Lisbon (SCECR Conference and Management Seminar), NYU Stern School of Business, Technis Webinar, Universitat Pompeu Fabra Barcelona, University of Zurich, Wharton School San Francisco Campus (Business and GenAI Workshop), and Weizenbaum Institute. We acknowledge support from the Swiss National Science Foundation for the project 100013\_197807.}}

% \author{\ }
\date{May 27, 2025}
\author[]{Christian Peukert*} 
\author[]{Florian Abeillon*} 
\author[]{J\'er\'emie Haese*} 
\author[]{\\ Franziska Kaiser**} 
\author[]{Alexander Staub*} 

\affil[]{*University of Lausanne, Faculty of Business and Economics (HEC)}
\affil[]{** University of Zurich}

\clearpage\maketitle
\thispagestyle{empty}

\newcommand{\Keywords}[1]{\par\noindent{\small{\em Keywords\/}: #1}}
%\newcommand{\JELclass}[1]{\par\noindent{\small{\em JEL classification\/}: #1} X} 
% marketing science & management science useskeywords only 

\vspace*{-2em}
\begin{abstract}
\onehalfspacing
\noindent
Artificial intelligence (AI) systems rely heavily on human-generated data, yet the people behind that data are often overlooked. Human behavior can play a major role in AI training datasets, be it in limiting access to existing works or in deciding which types of new works to create or whether to create any at all. We examine creators' behavioral change when their works become training data for commercial AI. Specifically, we focus on contributors on Unsplash, a popular stock image platform with about 6 million high-quality photos and illustrations. In the summer of 2020, Unsplash launched a research program and released a dataset of 25,000 images for commercial AI use. We study contributors' reactions, comparing contributors whose works were included in this dataset to contributors whose works were not. Our results suggest that treated contributors left the platform at a higher-than-usual rate and substantially slowed down the rate of new uploads. Professional photographers and more heavily affected users had a stronger reaction than amateurs and less affected users. We also show that affected users changed the variety and novelty of contributions to the platform, which can potentially lead to lower-quality AI outputs in the long run. Our findings highlight a critical trade-off: the drive to expand AI capabilities versus the incentives of those producing training data. We conclude with policy proposals, including dynamic compensation schemes and structured data markets, to realign incentives at the data frontier.
\vspace*{2em}
\Keywords{\footnotesize{Generative Artificial Intelligence, Training Data, Licensing, Copyright, Natural Experiment}}\vspace{3pt}

\end{abstract}

\newpage

\pagestyle{plain}
\setcounter{page}{1}

\doublespacing

%%%%%%%%%%%%%%%%%%%%%%%%
\section{Introduction}
%%%%%%%%%%%%%%%%%%%%%%%%

Data is an essential input to artificial intelligence (AI) which enables societal and economic progress through the discovery of knowledge, enhancement of productivity, and expansion into new or existing markets \citep{wu2020data,beraja2023data,mcelheran.2024.AIAdoptionAmerica, Brynjolfsson2025}. The rapid proliferation of generative AI (genAI) technologies has enabled hundreds of millions of users to generate high-quality text (including software code), images, audio, and video at a negligible cost. All this is made possible by the vast \textit{stock of data} publicly available on the internet, which, together with computing power, is responsible for the tremendous performance improvements in recent large language models \citep{ho2024algorithmic}. However, access to a continuous \textit{flow of data} remains crucial for performing a wide range of AI applications in dynamic social environments \citep{he2011incremental, valavi2022time,Peukert2024Editor}.
Input data for genAI models is primarily composed of online content produced by humans, in the form of text, images, sound, and video. Hence,  the incentives that drive contributions to training datasets, coupled with institutional frameworks, play a major role for the quality of genAI \citep{2023delRioChanona,burtch2024generative,Jia2025,Quinn2025}. 

Questions of whether, how, and which data can or should be used to train AI models have moved to the center of the policy debate.
Regulatory frameworks -- or the lack thereof -- can significantly impact both the \textit{supply} of and \textit{demand} for data. Concerning input data, AI policy intersects with privacy law and competition policy \citep{farboodi2019big,Jones2020,Azoulay2024}. More recently, however, intellectual property law has emerged as a critical area of the policy discourse because input data may be subject to copyright protection \citep{samuelson2023generative}. While some jurisdictions allow exemptions for research and development in copyright law, general legal uncertainty remains \citep{fiil2022legal,henderson2023foundation}. Several high-profile lawsuits against AI developers allege, among other claims, direct copyright infringement by creating unauthorized copies of their works and using such copies as training data.\footnote{See articles in the New York Times (\url{http://tiny.cc/pxljxz}) and Reuters (\url{http://tiny.cc/txljxz}).}
As a result, policymakers everywhere are tasked with balancing innovation in AI with the interests of rights holders \citep{appel2023generative,de2025intellectual}.

In this paper, we highlight how human behavior can shape the flow of data, studying the response of users to their works being made available for the training of commercial AI applications. The empirical setting for our study is Unsplash, one of the largest stock photography websites. In the summer of 2020, Unsplash released metadata on, and enabled the bulk download of, a subset of 25,000 images explicitly to train commercial AI applications (hereafter the ``LITE'' dataset). This natural experiment allows us to identify the causal effects of being included in AI training data and characterize the responses of those affected. We compare the upload behavior of contributors whose works were included in this dataset to contributors whose works were not. We then explore the underlying mechanisms by studying heterogeneity across contributors. Finally, we investigate changes in the variety and novelty of contributions in response to the release of the LITE dataset. We measure variety and novelty by calculating the similarity of each new upload to all existing images one year prior to the treatment based on textual descriptions of their content using natural language processing (NLP) methods. This allows us to track how variety and novelty change over time for both treated and untreated users, before and after the release of the LITE dataset.

We find that treated users left the platform at a higher-than-usual rate and, conditional on remaining active, substantially slowed down the rate of new uploads by about 40\% per month. Our analysis of mechanisms suggests that this reduction is driven by the economic interests of contributors. We show that professional photographers and prolific users reduce their contributions significantly more. In addition, the behavioral response appears to be stronger when AI is perceived as an economic threat. Users with multiple images in the training dataset reduce their activity almost twice as much compared to users who only had one image included. Second, the effect intensifies around two years later when capable genAI models become widely known to the public. We also provide evidence that users shift away from the type of photography that is most prominently featured in the LITE dataset.
However, our results do not suggest that photographers leave their profession altogether. For a subsample, we can show that users' behavioral changes occur only on Unsplash and are not mirrored in their activity on Instagram, the most popular image-sharing platform.

Concerning potential long-run effects for AI training data, 
our results illustrate that user behavior affects the size and the quality of the flow of data. 
Our content analysis shows that within-users, uploads decrease in variety but not in novelty compared to the existing stock of images. Across users, the variety of uploaded images decreased by about 5\% compared to the stock. Further, we find that affected users upload images that are about 30\% less novel. This shows that changes to the aggregated training dataset stem primarily from changes in user composition but also from a shift in individual behavior. Extrapolating the effects to the aggregate suggests that making the entire catalog available for commercial AI research (which is similar to a policy that would exempt data mining from copyright law) would have reduced the supply of data by half. At the same time, the data flow would become increasingly similar to the stock, with the number of very similar images tripling within a year. Relating our findings to experimental research in computer science, we conclude that our estimates of a decrease in the quantity, variety, and novelty of the flow of training data translate into a non-negligible decrease in AI output quality. 

We discuss managerial implications for platforms that face trade-offs between making data available for AI training and potential user responses, including measures to counteract decreased user activity. Furthermore, we outline policy proposals to address these negative consequences, including a dynamic compensation scheme for rights holders and data markets modeled on those used in targeted online advertising.

The contributions of this research are twofold. First, we provide a large-scale empirical study of user behavior in the context of AI training data. Thereby, we add to a discussion of how changes in the upstream supply of human-generated data can affect genAI model outputs, which remains an under-explored research topic \citep{Jia2025}. This supply-side perspective expands research on platform contributors' role in the AI economy \citep{2023delRioChanona, burtch2024generative, yiu2024ai, Quinn2025}, but also complements research on the demand-side, discussing the downstream effects of genAI on creative tasks \citep{Boussioux2024,Doshi2024,Zhou2024}, and the effects of AI on labor markets in general \citep{Felten2023,Eloundou2024,Hartley2024,Brynjolfsson2025,Demirci2025}. Second, our setting allows us to estimate the causal effects of adding human-made works to training datasets that can be used to develop commercial AI applications. Thereby, we provide empirical evidence to a primarily theoretical debate on policy-making in the age of AI \citep{fiil2022legal,henderson2023foundation,samuelson2023generative,Wang2024Economic,yang2024generative,gans2025copyright}.

%%%%%%%%%%%%%%%%%%%%%%%%
\section{Background and related literature}
%%%%%%%%%%%%%%%%%%%%%%%%

Regulation can affect the demand for data by defining rules about which data can or should be used for AI applications. 
For example, privacy law and competition policy can reduce firms' abilities to track consumer behavior \citep{johnson2022economic}, require firms to elicit consent from consumers \citep{godinhodematos.2022.ConsumerConsentFirm} or share data with competitors \citep{prufer2021competing,lei2023value}. Access to data has a direct impact on AI applications, causing the value of data -- especially for prediction tasks -- to depend on both its quality and quantity \citep{neumann2019,lei2023value,Peukert2024Editor,sun2023value}.

Concretely, the AI Act of the European Union (EU) demands that ``Training, validation and testing datasets shall be relevant, sufficiently representative, and to the best extent possible, free of errors and complete in view of the intended purpose.''\footnote{See Article 10(3) of the European Union Artificial Intelligence
Act signed in June 2024, \url{https://eur-lex.europa.eu/legal-content/EN/TXT/?uri=OJ:L_202401689\#art_10}.} However, such provisions also have intersections and perhaps even contradictions with intellectual property law. In the EU, for example, although copyright law was only recently modernized with the Directive on Copyright in the Digital Single Market in 2019, discussions about further reform are in full swing. When rights holders can exercise exclusive rights and restrict access to certain source materials, copyright may, in turn, create or promote biased AI systems \citep{levendowski2018copyright}. 

With the abundance of data online, researchers and AI developers can potentially access a large \textit{stock of data}, most of which was created before the dawn of AI. 
In general, surprisingly little is known about input datasets for AI applications. Recent empirical work demonstrates that AI models perform better when trained on larger and more diverse datasets, including potentially unauthorized sources. \cite{Jia2025} show that large language models exhibit significantly improved performance on books included in the Books3 dataset -- a collection of pirated books -- compared to those not present in Books3. Their findings highlight how restricted data access due to copyright laws could impact the development and accuracy of AI models. Other empirical contributions come from the computer science community. \cite{birhane2021large} study a widely-used image training dataset concerning privacy issues such as the presence of human faces, sexually explicit content, and measures of the age and gender distribution of depicted humans. Relatedly, \cite{guilbeault2024online} explore gender bias in about one million images downloaded from the web.

In many applications, access to a large stock of historical data is sufficient. Yet, in dynamic settings, where the stock of data can become outdated, a consistent \textit{flow of data} is key for algorithmic performance \citep{he2011incremental, valavi2022time, Peukert2024Editor}. However, the flow of data is endogenously determined by human behavior, which is, in turn, influenced by institutional settings such as privacy law, competition policy, and intellectual property law. This dynamic dimension is severely understudied.

Key criteria that define the quality of training datasets, be it in terms of stock or flow, concern the \textit{variety}  and \textit{novelty} of information in the dataset. To illustrate this, consider two examples. First, imagine a panel dataset that an econometrician would analyze. We can increase variety by adding more cross-sections, e.g., more firms, and we can increase novelty by expanding the time series, i.e., adding more years per firm. As a second example, consider a dataset with photos of New York City. We can add to the variety dimension by adding images of buildings from every neighborhood, from different angles and times of the day, etc. However, as some buildings make way for new developments, it becomes necessary to add novel images of the same concept, e.g., the skyline, to ensure a continued high-quality dataset. Thus, variety and novelty can be measured by changes in the flow of incoming data points and their relation to the existing stock of data. As a result, user behavior that impact the flow of data can also alter key dimensions of dataset quality and, therefore, the output quality of AI applications.

Recent theoretical work is a useful starting point to think about policy options to solve the trade-off between creator interests and AI innovation \citep{yang2024generative,gans2025copyright}.
Digital goods, such as user-generated content or creative works, tend to have high fixed costs and near-zero costs of copying. Intellectual property rights or alternative business models create rents that allow creators to recoup their fixed-cost investments. The proliferation of genAI systems suggests two potential outcomes: AI might reduce these monopoly rents, leading humans to produce only works with lower fixed costs, or AI might increase human productivity, effectively reducing fixed costs and requiring lower monopoly rents to incentivize creation. These effects can depend on factors such as the bargaining power of rights holders \citep{gans2025copyright} and the existing body of work in a particular domain \citep{yang2024generative}. A potential solution is a licensing mechanism that compensates rights holders with a royalty scheme that is determined by Shapley values of pieces of training data that end up in the final outcome \citep{Wang2024Economic}.

The behavior of contributors to training datasets, especially when large parts of the public internet are considered to constitute a training dataset, can have long-run effects on online communities. 
Evidence from platforms like Stack Overflow suggests that the introduction of genAI systems like ChatGPT can lead to reduced community participation, a decrease in the number of questions and users as well as a decline in the quantity and quality of answers \citep{burtch2024generative}, and an overall decrease in activity \citep{2023delRioChanona}. Three recent working papers are most directly related to our paper. \cite{Quinn2025} discuss how the quality of training data can change because of the reactions of users to the introduction of genAI. In the context of Stack Overflow, they show that while users reduce their contributions, the underlying quality of the contributions, and thus future training data, improves which has effects for the next iterations of AI models. \cite{huang.2023.GenerativeAIContentCreator} 
investigate the impact of genAI on content creators across two major Asian platforms, Lofter and Graffiti Kingdom, through quasi-experiments. The study reveals that introducing genAI tools on Lofter decreased creator activity, whereas banning these tools on Graffiti Kingdom had a mixed impact: The activity of non-churning creators increased, while that of contributors with copyright concerns decreased. Similarly, \cite{lin2024} studies the online art platform DeviantArt and shows that contributors decreased their publication volume by 22\% after the platform introduced a genAI feature that let users generate art with a click of a button. 

% Overall, ensuring a continuous supply of training data is a key challenge in platform governance. \cite{bhargava2020platform} illustrate how even small shifts in data regulation can reshape users' willingness to share information, with direct implications for downstream analytics or AI training.
In summary, while evidence on the impact of genAI trained on vast and intransparent datasets is growing, the lack of research on contributors' responses to the explicit use of their work for the development of commercial AI models forms the backdrop of our study.
 
%%%%%%%%%%%%%%%%%%%%%%%%
\section{Empirical setting, data and methods} %\section{Data and econometric model}
%%%%%%%%%%%%%%%%%%%%%%%%     

\subsection{The Unsplash research program}

Unsplash is a stock image platform hosting approximately 6 million photos and illustrations, submitted by roughly 360,000 contributors. According to Similarweb, Unsplash is among the top 5 most visited websites in the ``photography'' category globally, receiving about 30 million visits each month. Over the past decade, Unsplash has recorded a cumulative total of 1 trillion image views, coming from tens of millions of individual users and thousands of partner websites and applications.
Unlike other stock photography websites, the vast majority of images on Unsplash are made available under a permissive license, which allows them to be used freely for both non-commercial and commercial purposes. %This licensing model provides a clean empirical setting where we can identify behavioral responses to a new usage type -- nevertheless covered by the license -- which is exogenous to individual users. 
Compared to other websites that make images available under similar licenses (e.g., Wikimedia), contributions on Unsplash are of comparatively higher quality. Partially, this is enabled by rigorous moderation efforts of an editorial team that hand-selects images that are prominently shown on the website. As of May 2023, the curation team has selected about 300,000 images for curation, which represents about 6\% of all images available on the platform. 

% In 2023, Unsplash conducted a survey among contributors, which provides some descriptive statistics on the platform's user base.\footnote{See \url{https://unsplash.com/blog/unsplash-census-report-2023/}.} Among the 3,500 respondents, 67\% are under the age of 35, 70\% are male, and 17\% are based in the United States. The majority of users contributes photos (92\%), while a smaller fraction offers 3D renders (2.3\%), or both (5.7\%). Three-quarters of the contributors identify as amateurs and a third post images on a regular basis (every 2–-3 months).

In August 2020, Unsplash started a research program and released a dataset containing 25,000 images available for commercial and non-commercial use, including for the training of AI. This dataset is called the LITE dataset. In addition, the platform released a dataset covering all images hosted on the platform (hereafter the FULL dataset) made available exclusively for non-commercial research purposes. Access to the LITE dataset happens with the click of a button, while access to the FULL dataset is subject to an application process with a thorough manual review process.\footnote{By releasing the two flavors of the dataset, Unsplash essentially treated their entire community with their image (metadata) being available for AI research. However, the images that were selected for the LITE dataset receive a much stronger treatment because access is quick, free, and subject to less restrictive usage permissions. As we show in section \ref{sec:robust:controlgroups}, users who did not have images in the LITE dataset (and therefore only in the FULL dataset), do not seem to change the amount of contributions they make to Unsplash compared to their amount of contributions to Instagram. In contrast, we see a significant difference for users with images in the LITE dataset across Unsplash but not on Instagram. This implies that there is no causal change in behavior of being included in the FULL dataset without being part of LITE.} 

Both released datasets contain costly-to-compute metadata such as keyword tags, depicted landmarks, color distributions, etc., as well as the URLs to download the full-resolution images. The bulk download of the images themselves is explicitly permitted\footnote{In response to the question ``I would like to download
the images using the provided urls. Do I risk getting blocked for making too many
requests/downloads'', the Head of Data at Unsplash writes on GitHub: ``you shouldn't be
blocked if you use the provided URLs to download the photos. You can try it out and raise an issue if anything goes wrong.'' (See \url{https://github.com/unsplash/datasets/issues/37}.}, straightforward and fast (a sample script to download all images in the LITE dataset is provided in the appendix Listing \ref{lst:image_download}). As a result, the LITE release effectively provides easy access to both metadata and image files in a commonly accepted format for image training datasets.\footnote{Out of 38,679
datasets currently available on Huggingface, only 10,885 include image files - see \url{https://huggingface.co/datasets?modality=modality:image} and
\url{https://huggingface.co/datasets?modality=modality:image\&format=format:imagefolder}
} 

\subsubsection{The selection of images in the LITE dataset} \label{sec:emp:selection}

The LITE dataset was created on June 25, 2020, and, therefore, is a subset of all images available on Unsplash until that date. The images to be included in the dataset were selected based on four criteria: (1) images are nature-themed, (2) images have been curated by the editorial team, (3) images are currently available on the platform, (4) the total number of images in the dataset is maximum 25,000.\footnote{See \url{https://github.com/unsplash/datasets/issues/55} for details on the sampling procedure as expressed by Timothy Carbone, Head of Data at Unsplash, who runs the research dataset program. In an email conversation with us, he shared the database query he used to select the images in LITE (see Listing \ref{lst:query}) and clarified further: ``Focusing on a single keyword helps the end-user access more depth of content, rather than breadth. This is allegedly more helpful for things like training AI and will hopefully provide a better `trial' experience for the dataset.
Making sure we only choose curated content helps with lowering the risk of the content being removed from the platform and makes the dataset easier on the eye.''}
To determine whether an image is nature-themed, Unsplash uses computer vision technology that assigns keywords to images, most prominently Amazon's Rekognition service.\footnote{See \url{https://unsplash.com/blog/the-data-stack-at-unsplash/} and \url{https://docs.aws.amazon.com/rekognition/latest/APIReference/API_DetectLabels.html} for details.} These auto-generated keywords come with a confidence score, which ranges from 0 to 1. Curated images are selected by a human curator at Unsplash to be ``special'', meaning they possess a unique perspective, exploration of light and materials, creative use of color, inventive framing, mood evocation, or usefulness for reuse. These images were either featured on the front page of the website or featured for ten days under a specific topic (e.g.: ``nature'').\footnote{See \url{https://unsplash.com/blog/how-we-choose-what-photos-to-feature-on-the-unsplash-homepage/}.}

To pull the data for the LITE dataset, Unsplash used a query without specific ordering in the SQL query, i.e., no ORDER BY clause (see Listing \ref{lst:query}). In PostgreSQL, which is the database system they used at the time, this leads to an unpredictable row order of the results. The row order depends on the database's internal storage and retrieval mechanisms, which can be influenced by factors like the structure of the database's indexes and how the query optimizer chooses to execute the query.\footnote{The documentation of PostgreSQL, which is the database system Unsplash uses, states: ``When using LIMIT, it is a good idea to use an ORDER BY clause that constrains the result rows into a unique order. Otherwise, you will get an unpredictable subset of the query's rows [...] You don't know what ordering unless you specify ORDER BY. [...] If ORDER BY is not given, the rows are returned in whatever order the system finds fastest to produce.'', see \url{https://www.postgresql.org/docs/current/sql-select.html} and ``If sorting is not chosen, the rows will be returned in an unspecified order. The actual order in that case will depend on the scan and join plan types and the order on disk, but it must not be relied on. A particular output ordering can only be guaranteed if the sort step is explicitly chosen.'', see \url{https://www.postgresql.org/docs/current/queries-order.html}.} 

\begin{table}[!t]
 \caption{Sort order in LITE dataset} \label{tbl:order}
\input{Overleaf/tables/order}\\
%\noalign{\vskip .5em}
\begin{minipage}{\linewidth}
{\footnotesize
\def\sym#1{\ifmmode^{#1}\else\(^{#1}\)\fi}
\textbf{Note:} 
The dependent variable is the rank order of images in the LITE dataset. White-robust standard errors in parentheses. \sym{*} \(p<0.10\), \sym{**} \(p<0.05\) \sym{***} \(p<0.01\)
}
\end{minipage}
\end{table}

Indeed, Table \ref{tbl:order} shows that the order of images in the LITE dataset cannot be explained by image characteristics such as image popularity (i.e., views and downloads), whether the user has chosen the keyword ``nature'' to describe the image (\textit{UserKeyword}) or image age (i.e., time since upload).\footnote{Note that we do not have user-level information about 14 users because their accounts were deleted before we can observe them in FULL 1.0.1, the earliest FULL dataset we have access to. These users have 63 images in the LITE dataset. Hence, the number of observations in column 2 is smaller than in column 1.} Also, user characteristics such as the total sum of views and downloads of the users' images (\textit{UserPopularity}), the time since the first upload of the user (\textit{AccountAge}), and the total number of uploaded images before treatment (\textit{TotalUploads}) do not explain the order of images in the LITE dataset. Only the confidence score associated with the automatically assigned keyword ``nature'' (\textit{AutoKeywordScore}) is correlated with the sort order. The higher the confidence score of an image, the higher up it appears on the list. This is consistent with PostgreSQL using a descending order to generate a list of all keywords that meet the confidence condition. For example, conditional on meeting the threshold, keywords with a confidence score of 0.99 would appear before keywords with a confidence score of 0.90. However, it is worth stressing that the regression can only explain little variation in the rank order, as shown by the low value of the adjusted $R^2$.

With this information, we can construct a control group of images, for example, consisting of those images that would have appeared below row 25,000 had the query not used a LIMIT condition. Based on the results in Table \ref{tbl:order}, images would be listed further up, the higher the confidence value of the computer vision algorithm for the keyword ``nature''. For technical reasons, however, this does not necessarily mean that those images monotonically convey more of the concept ``nature''. They can also convey the same amount of ``nature'' but less of other concepts.\footnote{In multi-label classification algorithms using Deep Learning and Convolutional Neural Networks, as deployed in the DetectLabels service of Amazon Rekognition, the calculation of confidence scores is a nuanced process. While the exact technical underpinnings of Rekognition are not publicly disclosed, a slide deck from a System Dev Engineer at Amazon Webservices gives some indication about the service's architecture (see \url{https://iptc.org/download/events/phmdc2017/IPTC-PhMdC2017-AHornsby-AmazonRekognition.pdf}). The last layer typically applies the softmax function, transforming logits -- the network's final layer output, representing unnormalized log probabilities -- into a probability distribution across the predicted labels. Softmax is non-linear implying that changes in logits can be disproportionally reflected in confidence scores. Moreover, the confidence score for a given label is not only a function of its own logit but also depends on the logits of all other labels. Therefore, an increase in the confidence score of ``nature'', disproportionally so at the high end of the range, may result from a larger logit of the concept itself or from smaller logits of other concepts.}

\subsubsection{Announcement and stakeholders' awareness of the program}

The release of both research datasets was announced publicly on the Unsplash blog on August 6, 2020.\footnote{See \url{https://unsplash.com/blog/the-unsplash-dataset/}.} In this announcement, Unsplash mentioned the target users for the dataset: ``We're releasing the most complete high-quality open image dataset ever, free for anyone to use to further research in machine learning, image quality, search engines, and more.'' They also describe the differences between the two versions: ``We're releasing the data in two versions: a LITE dataset available for commercial and noncommercial usage, and the FULL dataset available for noncommercial usage.''.
Every user who subscribes to the Unsplash newsletter, which is the default option when signing up for an account, receives periodic emails summarizing announcements on the Unsplash blog. Therefore, we have reasons to believe that the average user was aware of the research dataset program. Users were not able to opt out, but were, of course, free to delete affected images or their user accounts. Users also did not receive monetary or other compensation for having their works included in the LITE or the FULL dataset.
The release of the Unsplash dataset was well received by the AI community. The GitHub repository, which provides documentation and sample code, was starred 2,500 times, and there is an active discussion with the community via issues and pull requests.\footnote{See \url{https://github.com/unsplash/datasets}.} In general, according to PapersWithCode, data from Unsplash has been used in at least 24 computer science papers for problems such as image augmentation, image generation, image inpainting and outpainting, and text-to-image generation.\footnote{See \url{https://paperswithcode.com/datasets?q=unsplash&v=lst&o=match}.}
Altogether, these are strong reasons to believe that the release of the LITE dataset was sufficiently salient to Unsplash users to allow us to detect treatment effects. Further, in section \ref{sec:res:mechanism}, we provide anecdotal evidence on the stated reactions of the photography community.

\subsection{Data and methods} \label{sec:data}

\subsubsection{Available information}

We have access to rich metadata on more than 4.9 million images, encompassing all images uploaded to Unsplash since May 2013 that remained available in either June 2020 or May 2023 (the FULL dataset). Specifically, we combine information from datasets released in August 2020 (but created on June 25, 2020) and May 2023.\footnote{To be precise, we have access to datasets in version 1.0.0 (LITE), and versions 1.1.0 and 1.2.1 (FULL). See \url{https://github.com/unsplash/datasets/blob/master/CHANGELOG.md}.} We observe the user account that has uploaded the image, the upload date, whether an image was curated, and a list of keywords associated with each image (both user-specified and generated by computer vision algorithms).
Using information about the camera gear that the photographer used, we flag users as professionals if at least one of the following conditions is fulfilled for any image they uploaded before the release of the LITE dataset. First, the exposure time is longer than 5 seconds, which cannot be executed by low-end cameras -- including most smartphones -- and without a tripod. Second, a photo was shot with a focal length of more than 250, which typically requires expensive professional lenses. Third, a photo is shot on a camera from a mid-to-high-end brand, i.e., not on a smartphone.\footnote{We do this based on the following list of brands: Hasselblad, Leica, Rollei, RED, Blackmagic, Voigtlander, Canon, Nikon, Fuji, Sigma, Olympus, Ricoh, Minolta, Konica, Yashica, Kodak, Agfa, Noritsu, Epson, DJI.} Another measure of whether a user is likely a professional photographer is whether they have chosen to display a badge on their Unsplash page that says ``Available for hire''. Finally, we know whether a user has joined the Unsplash+ program as a contributor (see section \ref{section:discussion:management} for more details on the program).\footnote{We do not know when a particular user joined the Unsplash+ program. However, we have this information as of April 23, 2024, approximately 2.5 years after the program's launch.}

\subsubsection{Treatment and control groups}

To analyze user-level contribution dynamics over time, we construct two separate panel datasets. First, although we lack high-frequency information on whether an image is still available on the platform, we have information on the upload date of every image from two snapshots that let us determine whether a user account with active images in June 2020 still has any images available in May 2023. Second, to study contribution patterns in detail, we aggregate data at the monthly level and count the number of uploads of every user within that time period. To create a balanced panel, we impute zeroes in months where we do not observe any activity.

To define treatment and control groups, we exploit the fact that Unsplash released a subset of their entire platform contents as a freely available dataset for commercial use. Because we want to study the behavioral responses of users when their work is made available as a training dataset for commercial AI, our treatment group is comprised of all 8,298 user accounts that have at least one image in the LITE dataset.\footnote{Users can change account names on Unsplash. Using the fact that the IDs of images remain unique and cannot be changed, we can track the same user even if they changed account names. The treatment concerns users with images present in the first release of the LITE dataset. Further, while de-facto randomization happened at the image level, mechanically, some users may have a higher chance of having images included in the LITE dataset because the distribution of contributions to the platform is not uniform. We correct for potential implications by estimating models with user-fixed effects that remove user-specific time-invariant variation (such as generally being more active on the platform), as well as by estimating models that weigh users inversely proportional to their pre-existing stock of contributions.} There are several ways to construct a control group. First, we could take all users who had uploaded at least one image before the LITE dataset was created but did not have any images included in the LITE dataset. Second, we could take the set of users whose images fulfilled the opposite of the selection criteria discussed in section \ref{sec:emp:selection}, i.e., users who had not uploaded nature-themed and curated images when the LITE dataset was created.
Third, we could take all users whose images fulfilled the same criteria as in the original database query but were not included in the LITE dataset. This approach comes closest to resembling the perfect counterfactual for users in the LITE dataset, making it our preferred specification. This specification of the control group includes 3,754 users.
In the robustness section, we also report results from the two former approaches, showing that the results are consistent, but our preferred approach yields the most conservative estimates.

Further, we can exploit the fact that some users contribute to Unsplash as well as Instagram, an unrelated image-sharing platform that is not affected by Unsplash's research program. 
Uploads to Instagram serve as a counterfactual of what upload behavior on Unsplash would have been had they not released the LITE and FULL datasets.
As we report in the robustness section, we show that only users whose data was released for commercial usage (i.e. were part of LITE) showed a causal reaction.

%%%%%%%%%%%%%%%%%%%%%%%%
\subsubsection{Identification strategy and econometric model}
%%%%%%%%%%%%%%%%%%%%%%%%     
%Identification
Our identification strategy is based on the notion that the images in the LITE dataset are randomly drawn from a superset of all images on the platform that meet the eligibility criteria.

For our econometric analysis, we rely on a difference-in-differences approach. We perform a set of OLS regressions saturated with fixed effects to compare users with images in the LITE dataset to users who had at least one image deemed eligible to be included in the LITE dataset but were not included in the LITE dataset.
Our baseline specification is
\begin{align}
    Y_{it}&=\delta (Post_t \times Treated_{i}) + \eta_t + \mu_i + \varepsilon_{it}, \label{eqn:diff-in-diff}
\end{align}

where $Y_{it}$ is the outcome of interest, e.g., the number of images uploaded by user $i$ in month $t$. $Treated_{i}$ indicates whether user $i$'s works were included in the LITE dataset. $Post_{t}$ is a binary variable that takes the value 1 if the current month is after the release of the LITE dataset, and 0 otherwise. We include month fixed effects $\eta_{t}$ in all specifications, and in some specifications, we additionally include user fixed effects $\mu_{i}$. User fixed effects capture any unobserved time-invariant variation across users, such as the higher likelihood of prolific contributors having their images included in the LITE dataset. The parameter $\delta$ estimates the causal effect of being included in a dataset made available for AI training on user behavior.

%%%%%%%%%%%%%%%%%%%%%%%%
\section{Results} 
%%%%%%%%%%%%%%%%%%%%%%%%
We begin by reporting baseline results separately for users' deletion rates as well as their uploading behavior. Then, we explore potential underlying mechanisms.

\subsection{Baseline results}

\subsubsection{Changes in deletion rates} 

First, we study whether being included in the LITE dataset affects the likelihood of an image or a user account being deleted from the platform. Hence, this set of results differs from all other analyses below, which focus on users' uploading behavior. 

It is important to note that while we know the exact upload date of each image, we do not know the exact deletion dates of either images or user accounts. However, we can use the fact that Unsplash periodically releases updates of the FULL dataset, which encompasses the entire platform catalog at a particular point in time. We can trace images (and corresponding users) that were part of the LITE dataset (created in June 2020) and observe whether they were still available on the platform in May 2023.

\begin{table}[!t]
 \caption{Results: Likelihood of remaining on the platform} \label{tbl:survival}
\input{Overleaf/tables/new/survival_image_user}\\
%\noalign{\vskip .5em}
\begin{minipage}{\linewidth}
{\footnotesize
\def\sym#1{\ifmmode^{#1}\else\(^{#1}\)\fi}
\textbf{Note:} We estimate a linear probability model. The dependent variable is an indicator for whether an image (columns 1-3) or a user account (columns 4-5) remains on Unsplash as of May 2023. In column (1), the sample includes all images. In column (2), the sample is restricted to images of user accounts with at least two images in the LITE dataset. In column (3), we add user fixed effects. In columns (4) and (5), the outcome switches to the user level (i.e., whether a user account is still active). Column (4) includes all users, and column (5) again restricts to users with at least two images in the LITE dataset. The variable \emph{Treated} indicates whether a user was part of the LITE dataset. We compare survival rates over 2.5 years (from Aug 2020 to May 2023).  We include fixed effects for the image's age (in months) in all specifications. Standard errors are clustered at the user level in parentheses. \sym{*} \(p<0.10\), \sym{**} \(p<0.05\) \sym{***} \(p<0.01\)
}
\end{minipage}
\end{table}

In column (1) of Table \ref{tbl:survival}, we see that images that were included in the LITE dataset (i.e., that were treated) have a 2 percentage points higher chance of surviving on the platform than images that were not included in the LITE dataset.\footnote{Calculated as the coefficient (column 1: 0.0220) divided by the mean survival rate of the control group (columns 1-3: 0.9075)} In column (2), we limit the sample to works of photographers that have at least two images included in the LITE dataset and find the same result. We do so to compare against the specification in column (3), which has user-fixed effects. The coefficient there implies a 1 percentage point higher likelihood of survival. This suggests substantial differences in the variation between users and within users. Overall the results suggest that treated images are deleted at a rate of 7-8\% ($\approx$ 3\% per year), whereas control images are deleted at a rate of 9\% ($\approx$ 4\% per year)\footnote{For the control group, the deletion rate is calculated as 1 minus the mean survival rate of the control group (columns 1-3: 0.9075) while the coefficient (column 1: 0.0220) is added to the control group's mean survival rate before subtracting to arrive at the treated group's deletion rate. The yearly approximate deletion rate is derived by averaging across the time available in our sample (2.5 years).}. The observed effects on deletion likely stem from substantial heterogeneity in user responses, consistent with anecdotal evidence we discuss in Section \ref{sec:res:mechanism}. We provide quantitative evidence supporting this heterogeneity in Table \ref{tbl:survival_image_usertypes}. Specifically, users investing in expensive camera gear are significantly less likely to maintain their images on the platform (significant in column 3), aligning with qualitative evidence suggesting professional photographers are particularly critical of their data use for AI. Conversely, images belonging to users who subsequently display a "For Hire" show increased persistence on Unsplash, indicating perceived business opportunities outweigh potential AI displacement concerns. Furthermore, images from users experiencing more intense treatment (those with multiple images included in the LITE dataset) exhibit higher deletion likelihood, supporting the argument presented later (Section \ref{sec:res:mechanism:ai}) that more intensely treated users significantly reduce their future upload activity.

In columns (4) and (5), we turn to user accounts. We see that treated user accounts are less likely to survive than accounts of control users. The estimate is about 1 percentage point in both samples of all users and those users with at least two observations (i.e., the users in the sample in columns 2 and 3).
This implies that accounts of treated users are deleted at a rate of 3\% ($\approx$ 1\% per year), whereas control users delete their accounts at a rate of 4\% ($\approx$ 1.5\% per year).

%%%%%%%%%%%%%%%%%%%%%%%%
\subsubsection{Changes in users' uploading behavior}\label{sec:res:uploads}
%%%%%%%%%%%%%%%%%%%%%%%%

\begin{figure}[!t]
\caption{Number of uploads, treatment versus control group} \label{fig:diff_number_photos}
\begin{minipage}{.5\linewidth}
\footnotesize \textit{A: Levels}\\
        \includegraphics[trim=10 0 0 0 , clip, width=\textwidth]{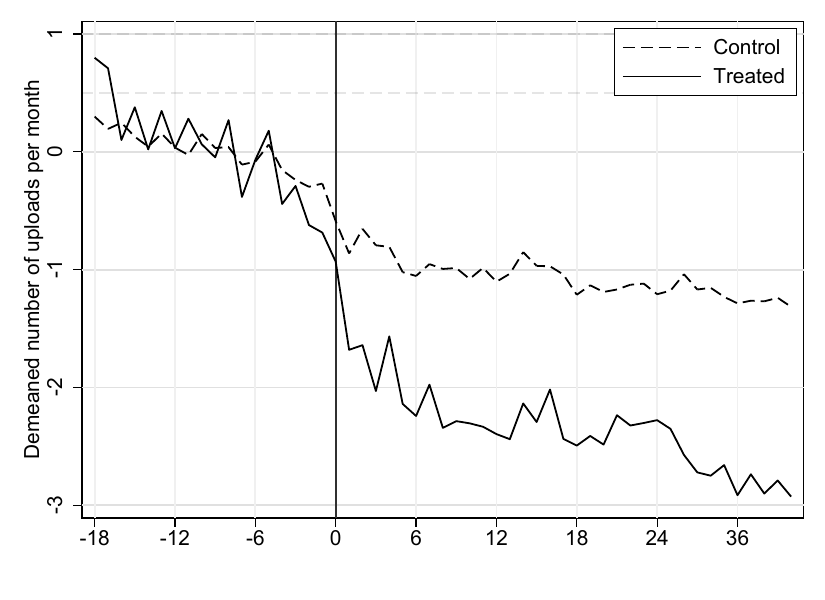}\\
\end{minipage}
\begin{minipage}{.5\linewidth}
\footnotesize \textit{B: Differences}\\
        \includegraphics[trim=10 0 0 0 , clip, width=\textwidth]{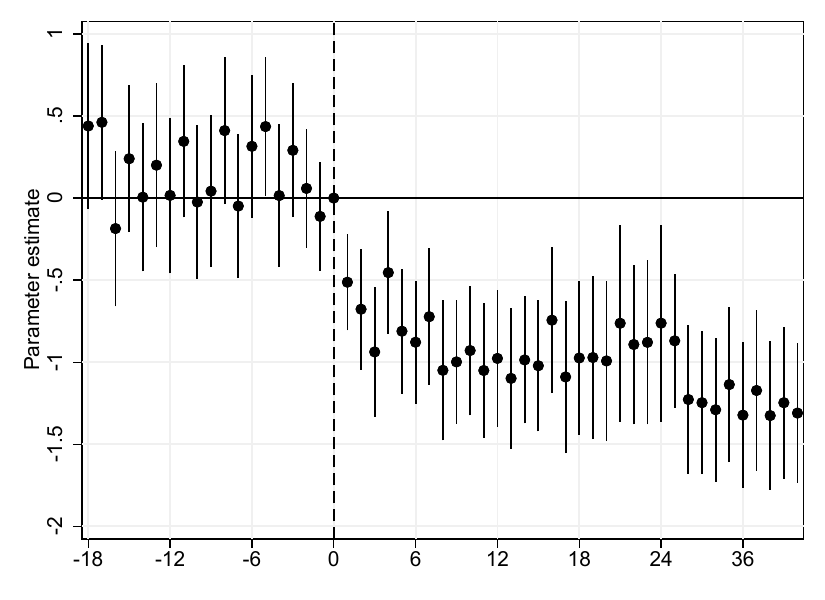}\\
\end{minipage}\\
{\footnotesize
\textbf{Note:} In Panel A, we depict the demeaned average number of uploaded images for users in the treatment group (solid) and users in the control group (dashed). We use the group-specific pre-period mean number of uploaded images to demean (3.44 for the treatment group and 1.33 for the control group). In Panel B, we plot OLS estimates of the $\delta_{\tau}$ coefficients obtained from estimating $Y_{it} = \sum_{\tau \in T} \nu_{\tau} + \sum_{\tau \in T} \delta_{\tau} \left( \gamma_{\tau} \times  Treated_{i} \right) + \mu_{i} + \varepsilon_{it}$, with the number of uploads of user $i$ per month $t$ as the dependent variable. The dots reflect month-specific point estimates comparing treated users to the control group. Standard errors are clustered at the user level, and bars indicate 90\% confidence bands.
In both panels, the vertical line indicates the creation of LITE 1.0.0.
}
\end{figure}

We now focus on the number of new uploads as the dependent variable. We compare users with at least one image in the LITE dataset to the control group. For this analysis, we can draw upon high-frequency information, which allows us to construct a panel dataset at the user-month level. In Panel A of Figure \ref{fig:diff_number_photos}, we plot the average number of uploads per month in the treatment group and the control group. For ease of interpretation, we demean using group-specific pre-period values. In the observed 18 months before the creation of the LITE dataset (indicated by zero on the horizontal axis), upload rates do not fluctuate much for both. Looking at differences between treated and control in Panel B of Figure \ref{fig:diff_number_photos} confirms that there are no statistically significant differences in the before period (all month-specific differences are jointly not significantly different from zero). The lack of statistically significant deviations from 0 in the pre-period supports the parallel trends assumption of the difference-in-differences model. After the policy change is communicated, we see a substantial and statistically significant decrease in the after period. The effect is immediate and remains relatively stable over time. After 24 months, there seems to be a further decrease, which we investigate further in section \ref{sec:res:mechanism:ai}. 

\begin{table}[!t]
 \caption{Results: Changes in uploads} \label{tbl:uploads}
\input{Overleaf/tables/new/uploads_fe}\\
%\noalign{\vskip .5em}
\begin{minipage}{\linewidth}
{\footnotesize
\def\sym#1{\ifmmode^{#1}\else\(^{#1}\)\fi}
\textbf{Note:} 
The dependent variable in columns (1) and (2) is the total number of uploads per month, in columns (3) and (4) it is an indicator of whether a user account has uploaded at least one image in a given month, and it is the log number of uploads in columns (5) and (6). \textit{Treated} indicates whether a user was part of the LITE dataset. All specifications include fixed effects for the upload month of the image; the specifications in columns (2), (4), and (6) additionally include user-fixed effects. Standard errors are clustered at the user level in parentheses. \sym{*} \(p<0.10\), \sym{**} \(p<0.05\) \sym{***} \(p<0.01\)
}
\end{minipage}
\end{table}

We can quantify the average treatment effect for the entire observed post-period in an OLS model specified in equation (\ref{eqn:diff-in-diff}). The results are reported in Table \ref{tbl:uploads}. In column (1), we look at the number of uploaded images per month and quantify the average treatment effect as 1 less upload per month, which translates into a reduction of about 37\%. In column (2) we add user-fixed effects to capture any user-specific time-invariant variation. The results remain similar and statistically indistinguishable. Further, in columns (3) and (4), we document a decrease in the extensive margin of about 30\%, i.e., when we use a dependent variable that indicates whether a user has uploaded at least one image in a given month. Finally, in columns (5) and (6), we get similar results when we apply a log transformation to the dependent variable. 

\subsubsection{Extrapolation to aggregate effects}

We can now use the average treatment effect estimates to extrapolate aggregate and long-run effects to effectively simulate counterfactuals by varying the proportion of treated users. To do this, we define three key variables: $\lambda$, the share of users in the LITE dataset; $\hat{\delta}$, the effect of inclusion in LITE; and $g$, the factual post-period growth rate in images per user. The counterfactual growth rate, had LITE not been released, is $\bar{g} = g / (1 - \lambda + \lambda \hat{\delta})$. Using estimates from column (2) of Table \ref{tbl:uploads}, we calculate $\hat{\delta}=1-(1.1099/2.9038)=0.6$. When the LITE dataset was created in June 2020, Unsplash had a stock of $S_0=1,931,324$ images uploaded by $U_0=198,792$ users. 
The LITE dataset encompasses $U_L=8,298$ users, such that the share of users in the LITE dataset was $\lambda=U_L/U_0=0.042$.
About three years later, in May 2023, Unsplash had $S_t=4,960,120$ images and $U_t=338,635$ users. Hence, the factual growth rate $g=\frac{S_t}{U_t}/\frac{S_0}{U_0}-1=14.65/9.72-1=0.51$ and therefore the counterfactual growth rate $\bar{g}=(0.51)/(1-0.042+0.042\times0.6)=0.52$.
In absolute terms, without the release of the LITE dataset, users would have uploaded 19,313 more images to Unsplash ($=S_0(0.52-0.51)$).
We can now extrapolate that a twice-as-large LITE dataset would have reduced the flow of data by 2\% (=1-0.50/0.51) compared to the factual while making the entire catalog available for commercial AI research would have reduced the flow of data by 39\% (=1-0.31/0.51), corresponding to 386,264 fewer images uploaded over a three-year period.

\subsubsection{Robustness}

\paragraph{Probability weighting}

Table \ref{tbl:uploads_weighted} provides a version of the baseline specification (with month and user fixed effects) but additionally weights observations inversely by the number of images the user uploaded to the platform before the LITE dataset was created. This accounts for a mechanically higher probability of having images included in the LITE dataset if a user is more prolific. The results remain very similar and statistically indistinguishable from the baseline results (columns (2), (4), and (6) in Table \ref{tbl:uploads}) in the sense that 90\% confidence bands of the coefficients overlap.

\paragraph{Alternative control group definitions} \label{sec:robust:controlgroups}

%within Unsplash
Table \ref{tbl:uploads_alt_control} uses the same specification as in our baseline estimations in Table \ref{tbl:uploads} but with different definitions of the control group. In columns (1)--(3), we report results using a control group composed of users that did not have nature-themed (i.e., not with the keyword ``nature'') and no curated images at the time when the LITE dataset was created. The effect size for the number of uploads is 47\% in column (1), 33\% in terms of the extensive margin in column (2), and 19\% when using a specification with the log-transformed number of uploads in column (3).
In columns (4)--(6), we use a control group comprised of all users whose images were not included in the LITE dataset. The effect sizes are 50\%, 40\%, and 21\%, respectively. Note that the estimates reported in Table \ref{tbl:uploads_alt_control} are not significantly different from those reported in Table \ref{tbl:uploads} in that 90\% confidence bands overlap.

\paragraph{Comparison to upload behavior on Instagram} \label{sec:robust:insta}

\begin{figure}[!t]
\caption{Number of uploads, Instagram vs. Unsplash} \label{fig:diff_number_photos_insta}
\begin{minipage}{.33\linewidth}
\footnotesize \textit{A: Levels}\\
        \includegraphics[trim=10 0 0 0 , clip, width=\textwidth]{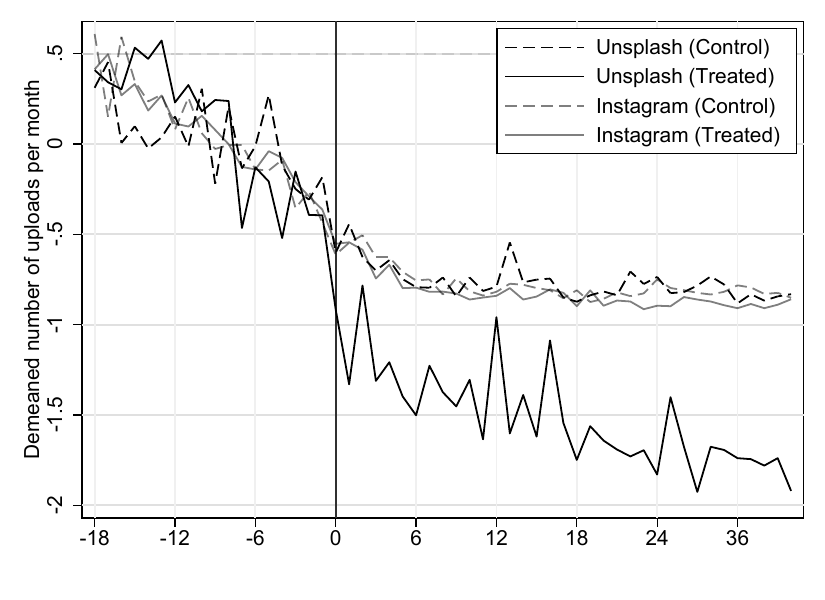}\\
\end{minipage}
\begin{minipage}{.33\linewidth}
\footnotesize \textit{B: Insta-Unsplash diff within control}\\
        \includegraphics[trim=10 0 0 0 , clip, width=.95\textwidth]{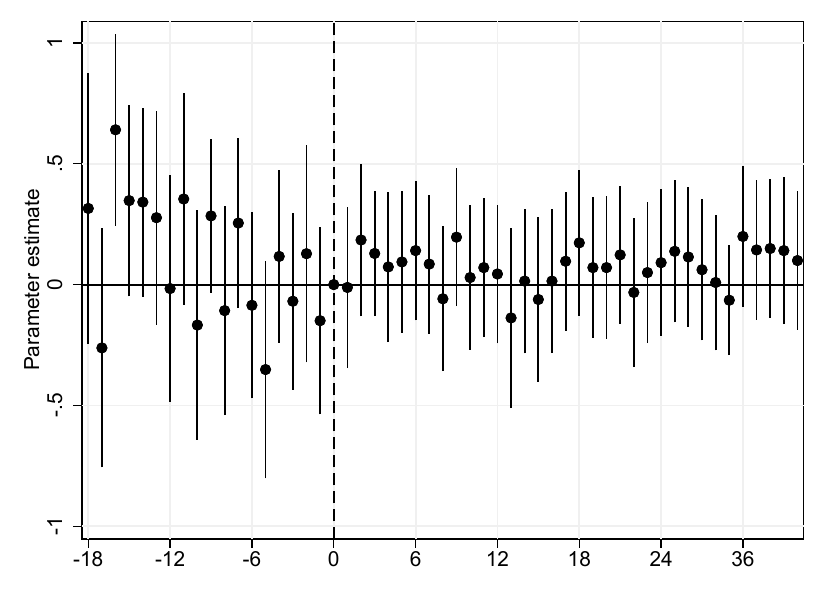}\\
\end{minipage}
\begin{minipage}{.33\linewidth}
\footnotesize \textit{C: Insta-Unsplash diff within treatment}\\
        \includegraphics[trim=10 0 0 0 , clip, width=.95\textwidth]{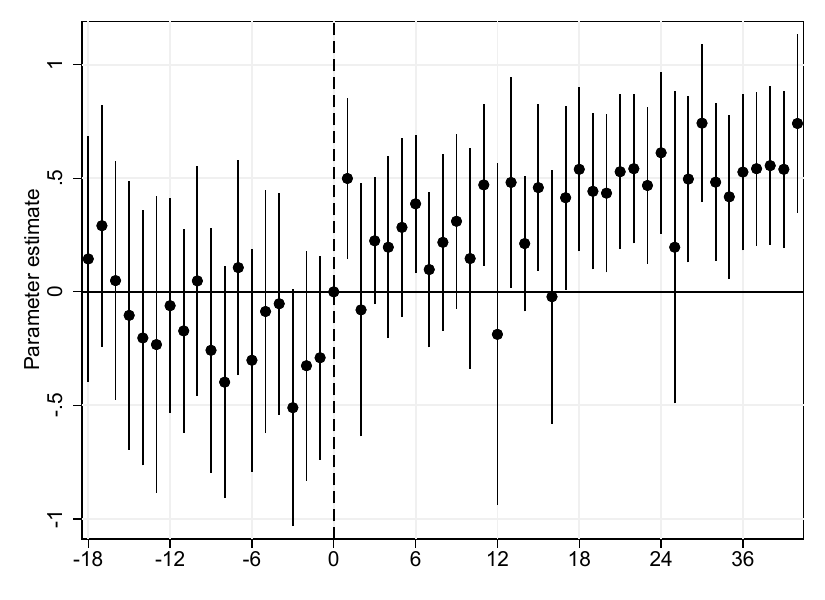}\\
\end{minipage}
\\
{\footnotesize
\textbf{Note:} In Panel A, we depict the demeaned average number of uploaded images for users in the treatment group (solid) and users in the control group (dashed). Black lines indicate uploads to Unsplash, whereas grey lines indicate uploads to Instagram. We use the group-specific pre-period mean number of uploaded images to demean (2.00 [0.97] for the treatment group on Unsplash [Instagram] and 0.81 [0.90] for the control group on Unsplash [Instagram]). In Panels B and C, we plot OLS estimates of the $\delta_{\tau}$ coefficients obtained from estimating $Y_{ijt} = \sum_{\tau \in T} \nu_{\tau} + \sum_{\tau \in T} \delta_{\tau} \left( \gamma_{\tau} \times  Instagram_{i} \right) + \mu_{i} + \varepsilon_{ijt}$, where $j$ indicates the treatment/control group with the number of uploads of user $i$ per month $t$ as the dependent variable. In Panel B (C), we report results from estimating this model on the subset of users in the Unsplash control group (treatment group). The dots reflect month-specific point estimates comparing upload behavior on Instagram to upload behavior on Unsplash. Standard errors are clustered at the user level, and bars indicate 90\% confidence bands.
In all panels, the vertical line indicates the creation of LITE 1.0.0.
}
\end{figure}

%using Instagram
One could speculate that the Unsplash research program might influence overall upload behavior on Unsplash, not only among users whose images appear in the LITE dataset (available for commercial applications) but also among those whose images are only included in the FULL dataset (available only for noncommercial research). To test this, we leverage the fact that a subset of Unsplash users is also active on the image-sharing platform Instagram. This allows us to compare the upload behavior of the same user (either with images in the LITE dataset or not) on Unsplash to their upload behavior on Instagram, which is not affected by Unsplash's research program. 
Especially for professional users, Unsplash and Instagram serve as comparable platforms for building an audience, and both require contributors to grant the platform a similar license when publishing images.\footnote{According to Instagram's terms of service, users retain ownership of any intellectual property rights that they hold in the content. However, by posting to Instagram, users grant the platform a non-exclusive, royalty-free, transferable, sublicensable, worldwide license to use the content they post.
Unsplash operates in a similar, yet more permissive fashion. When a user uploads an image to Unsplash, they grant anyone a license to use their image for free, including for commercial purposes, without requiring permission from or providing attribution to the user. Instagram's terms are not explicitly clear about commercial use by third parties unless it is sublicensed, while Unsplash explicitly allows commercial use by anyone. See \url{https://help.instagram.com/478745558852511/} and \url{https://unsplash.com/license}.} For a subset of 1,084 users who disclosed this information in their Unsplash profiles, we have their Instagram account names and have gathered data on their upload history on Instagram.\footnote{
Instagram account information is not available in the LITE or FULL datasets. We, therefore, do not have time-varying information, but we were able to get a snapshot on February 5, 2024, from which we observe Instagram accounts and then collected information about their Instagram activities. Further, technical restrictions require us to limit ourselves to the 20 most recent Instagram posts of a user. We exclude users for whom we only observe posts in the after period as we do not know whether this is because the account did not exist in the before period, whether there were no posts in the before period, or whether the user is so active that we can only see posts from the after period in our snapshot of the 20 most recent posts.}
Panel A of Figure \ref{fig:diff_number_photos_insta} shows the demeaned average number of monthly uploads to Unsplash (black) and Instagram (grey). We distinguish between uploads of users who have images in the LITE dataset (``treated'', depicted with solid black and solid grey lines) and uploads of users in our preferred control group (i.e. similar types of images but they are only part of the FULL data, depicted with dashed black and grey lines). The plot suggests that the upload behavior of the control group does not differ between Instagram and Unsplash. This is confirmed in Panel B, which shows insignificant coefficients from a regression that tests, within users that are part of the Unsplash control group, whether the monthly number of uploads differs between Instagram and Unsplash.

\paragraph{OLS versus non-linear models}

Given that our outcome variable is both a count, i.e., a non-negative integer, and zero in 86.6\% of the observations, one might be worried that OLS is not the optimal choice of estimator \citep{wooldridge2023simple}. Table \ref{tbl:uploads_user_percentiles_ppml} shows that our baseline results are broadly robust to using a Poisson pseudo-maximum likelihood regression (PPML) approach.\footnote{We do not report results from a negative binomial model because PPML is preferred for its robustness \citep{blackburn2015relative}.} However, the results indicate that PPML is sensitive to large values of the DV, which is of economic interest in our setting (i.e., prolific contributors). In addition, the interaction effect in non-linear models cannot be directly interpreted as the average treatment effect on the treated \citep{puhani2012treatment} while simultaneously imposing additional restrictions for parallel trends to hold \citep{roth2023when}. For these reasons, and simplicity in the interpretation of effect sizes, we report estimates from OLS regressions throughout.

\subsection{Mechanisms}\label{sec:res:mechanism}

In this section, we examine the potential underlying mechanisms for the baseline results. First, we provide anecdotal evidence highlighting concerns from the photography community, which we use to inform a heterogeneity analysis of the observational data. In particular, we analyze the economic motives of contributors, assessing whether professional and prolific contributors respond differently. We also examine whether the degree of exposure to the LITE dataset and growing awareness of the potential of genAI models influences user behavior. Finally, we investigate whether affected users alter the type of images they upload and whether they also change their behavior on another image-sharing platform or disengage from the profession entirely.

\subsubsection{Anecdotal evidence}

Discussions in online forums and blog posts offer insights into the reasons for contributors' negative reactions to the Unsplash research program. Comments reflect a polarized response, with strong opinions on its impact on the photography industry. The predominant sentiment is negative, particularly among professional photographers who argue that Unsplash devalues their work and undermines the sustainability of their profession. Many describe the platform's research program as exploitative, asserting that it effectively makes photographers give away their work for ``exposure'' while Unsplash may profit from partnerships with major companies. Others highlight how the program will contribute to the decline of traditional stock photography, making it increasingly difficult for photographers to earn a living through image licensing. Some commenters provide neutral, clarifying remarks about the dataset's content by correcting misconceptions regarding its licensing terms and explaining that it contains metadata rather than actual image files. A minority of users express support for Unsplash, viewing it as a valuable resource for creative and research purposes, particularly for those who do not rely on photography for income. The following quotes illustrate the overall sentiment:

\small
\begin{quote}
``If food could be obtained for free, the supermarkets would not be in business too long. Why would photographers regularly give away their licenses and claims to exclusivity to an outfit that will just parcel out their products for free? [\ldots]'' (Eric on \href{https://petapixel.com/2020/08/07/unsplash-releases-2-million-images-as-massive-open-source-dataset/}{Petapixel.com})
\end{quote}

\begin{quote}
``[\ldots]\ unsplash perhpas isnt for professionals, it is for hobbiest, and semipro people who need to showcase their work. for me free for all is great way to publish my work and im just a semihobbiest ;)'' (survivor303 in response to Eric on \href{https://petapixel.com/2020/08/07/unsplash-releases-2-million-images-as-massive-open-source-dataset/}{Petapixel.com})
\end{quote}

\begin{quote}
`` [\ldots]\ This [Unsplash's release of datasets] isn't part of the evolution of digital photography as some have said. It's just predatory. [\ldots]'' (Lahmajunior on \href{https://www.dpreview.com/news/2164828014/unsplash-releases-massive-open-source-image-dataset-with-2m-high-quality-photos}{DPReview.com})
\end{quote}

\begin{quote}
``This [Unsplash's release of datasets] is just the normal progression of a commoditized profession... Sad, but it is the reality of photography. It is a vicious cycle, when photographers cannot make money, they don't buy new gear. [\ldots]'' (KC on \href{https://petapixel.com/2020/08/07/unsplash-releases-2-million-images-as-massive-open-source-dataset/}{Petapixel.com})
\end{quote}

\begin{quote}
``[\ldots]\ professional photography will or is already going the same path as countless other professions went before in history. Namely that the know-how of professional is getting somehow devalued by new technology, which allows for cheap mass production of a good instead of labour intensive craftsmanship. [\ldots]'' (Toni Shalmonelli on \href{https://www.dpreview.com/news/2164828014/unsplash-releases-massive-open-source-image-dataset-with-2m-high-quality-photos}{DPReview.com})
\end{quote}
\normalsize

We distill two key insights: First, the primarily negative reactions to Unsplash's research program by the group of users that contribute images to the platform (i.e., photographers) seem very much in line with our baseline results. Second, the response may not be uniform but may be stronger among contributors who have more to lose from their work being freely available for AI training purposes (i.e., professionals).

\subsubsection{Economic motives}

Building on anecdotal evidence, we systematically test whether user behavior on Unsplash differs based on economic motives. First, we distinguish between professional and amateur contributors.
The results in column (1) of Table \ref{tbl:uploads_user_types} show that treated users who upload images taken with professional gear react even more negatively than others. Similarly, we find in column (2) that treated users with a badge on their profile page that says ``Available for hire'' reduced their uploads disproportionately compared to users without a badge. 
It is possible that users with professional gear and those who signal that they are available for hire make a living as photographers and are, therefore, perhaps more directly affected by the potential displacement effects of AI. Hence, monetary motives may be one of the drivers for users to contribute less to the platform after their works have been made available for AI research.

\begin{table}[!t]
 \caption{Results: Changes in uploads, professionals versus amateurs} \label{tbl:uploads_user_types}
\input{Overleaf/tables/new/uploads_user_types}\\
%\noalign{\vskip .5em}
\begin{minipage}{\linewidth}
{\footnotesize
\def\sym#1{\ifmmode^{#1}\else\(^{#1}\)\fi}
\textbf{Note:} 
The dependent variable is the number of uploads in a given month. \textit{Treated} indicates whether a user was part of the LITE dataset. \textit{Pro-Gear} indicates whether the user has uploaded at least one photo taken with professional gear (see section \ref{sec:data} for a detailed definition). \textit{ForHire} indicates whether a user has chosen to show ``Available for hire'' on their Unsplash profile page. The number of observations differs from those in Table \ref{tbl:uploads} because we do not observe \textit{Pro-Gear} for all images and \textit{ForHire} for all users. Fixed effects for upload month of the image and user account. Standard errors are clustered at the user level in parentheses. \sym{*} \(p<0.10\), \sym{**} \(p<0.05\) \sym{***} \(p<0.01\)  \sym{*} \(p<0.10\), \sym{**} \(p<0.05\) \sym{***} \(p<0.01\)
}
\end{minipage}
\end{table}

Further, it is worth noting that we observe that users in the treatment group are three times more likely to be contributors in the Unsplash+ program (t-statistic in a two-sided t-test is 6.13). This program enables users to gain monetary compensation and explicitly disallows the use of their works for AI training.\footnote{See \url{https://unsplash.com/plus/license} and \url{https://unsplash.com/blog/contribute-to-unsplash/}.} Hence, we interpret the substantially higher uptake of Unsplash+ among users who had at least one image in the LITE dataset as a response to counteract economic forces and to avoid the utilization of their works in AI applications.

Another, possibly correlated dimension of user heterogeneity concerns upload activity, which is an important metric for gaining insights into platform governance. In Table \ref{tbl:uploads_user_percentiles}, we show that the main effect reported in column (1) of Table \ref{tbl:uploads} varies by percentile of a user's total upload activity. The point estimates in columns (1)--(4) of Table \ref{tbl:uploads_user_percentiles} are smaller and not always significantly different from the estimate in column (1) of Table \ref{tbl:uploads} in the sense that 90\% confidence bands overlap. However, in absolute terms, the implied effect sizes are larger (column 2: -45\%, column 3: -44\%, and column 4: -40\%) compared to the -30\% of the baseline. Given that we observe the universe of activity on Unsplash, we prefer not to remove particularly prolific users and report estimates of average treatment effects based on all users as our more conservative baseline results.

\subsubsection{Treatment intensity} \label{sec:res:mechanism:ai}

\begin{table}[!th]
 \caption{Results: Changes in uploads, treatment intensity} \label{tbl:uploads_mechanisms}
 {\small
\input{Overleaf/tables/new/uploads_mechanisms}\\
}
%\noalign{\vskip .5em}
\begin{minipage}{\linewidth}
{\footnotesize
\def\sym#1{\ifmmode^{#1}\else\(^{#1}\)\fi}
\textbf{Note:} 
The dependent variable is the total number of uploads per month. \textit{Post} indicates the entire period after the creation of LITE.  \textit{TreatedSingle} indicates whether a user had exactly one image in the LITE dataset. \textit{TreatedMultiple} indicates whether a user had more than one image in the LITE dataset. \textit{Treated} indicates whether a user was part of the LITE dataset. \textit{PostAug22} indicates the period after August 2022, when groundbreaking genAI for images was released.   Fixed effects for upload month of the image and user account. Standard errors are clustered at the user level in parentheses. \sym{*} \(p<0 .10\), \sym{**} \(p<0.05\) \sym{***} \(p<0.01\)
}
\end{minipage}
\end{table}

The results in Table \ref{tbl:uploads_mechanisms} provide more evidence of the potential mechanisms through which the inclusion in the LITE dataset induces users to reduce their upload activity.

In column (1), we distinguish between users that were affected more or less intensively by the release of the LITE dataset. In particular, we distinguish between users who had one image included (about 60\%) and users who had multiple images included (about 40\%). Conditional on more than one image in the LITE dataset, the average number of images per user is 5.96 (standard deviation 14.18 and maximum 419).
The estimated effect size is 22\% for users with only one image included versus 45\% for users with multiple images.. This difference is statistically significant. These results provide further evidence that users must have been aware of their images being included in the LITE dataset because otherwise, we would not expect a stronger response from more heavily treated users. 

In column (2), informed by Figure \ref{fig:diff_number_photos}, we split the post-period into two parts. The first 24 months correspond to the period between June 2020 and August 2022, which garnered modest public interest in AI and AI art in particular. After August 2022, however, there was a substantial increase in media coverage of the topic, e.g., a New York Times article titled ``We Need to Talk About How Good A.I. Is Getting'', which may have further raised awareness about genAI in the Unsplash community.\footnote{See \url{https://www.nytimes.com/2022/08/24/technology/ai-technology-progress.html}} The timing also coincides with the public release of Stable Diffusion, the first text-to-image generator that could be run on consumer-grade machines, democratizing access to genAI technology after similar technology released a few months earlier had only restricted public access (e.g., DALL-E and Midjourney).\footnote{See \url{https://stability.ai/news/stable-diffusion-announcement}.} 
We find a treatment effect of 34\% before August 2022 and 49\% thereafter, which corresponds to a significant 40\% increase in the treatment effect. This suggests that, as awareness of the commercial potential of image generation AI rises, photographers reduce their contributions to Unsplash and, therefore, to the flow of AI training data.

\subsubsection{Differentiation or exit?} \label{sec:res:exit}

\begin{table}[!t]
 \caption{Results: Changes in uploads, by types of images} \label{tbl:uploads_image_types}
\input{Overleaf/tables/new/uploads_image_types}\\
%\noalign{\vskip .5em}
\begin{minipage}{\linewidth}
{\footnotesize
\def\sym#1{\ifmmode^{#1}\else\(^{#1}\)\fi}
\textbf{Note:} 
The dependent variable is an indicator of whether a user account has uploaded at least one image in a given month. Column (1) includes only Nature-themed images, column (2) includes only Curated images, and column (3) focuses on images that are both Nature-themed and Curated.\textit{Treated} indicates whether a user was part of the LITE dataset. Fixed effects for upload month of the image and user account. Standard errors are clustered at the user level in parentheses. \sym{*} \(p<0.10\), \sym{**} \(p<0.05\) \sym{***} \(p<0.01\)
}
\end{minipage}
\end{table}

Next, we study whether contributors differentiate and upload different types of images to Unsplash. In column (1) of Table \ref{tbl:uploads_image_types}, we show that the likelihood of uploading at least one nature-themed image decreases by 23 percentage points or 30\% relative to the average in the before period. Note that this resembles approximately the same effect size we get when looking at all types of images in column (2) of Table \ref{tbl:uploads}. However, columns (2) and (3) of Table \ref{tbl:uploads_image_types} suggest that the extensive margin decreases at 60\% for curated images and at 70\% for nature-themed and curated images. If we believe that Unsplash's curation team picks high-quality works, then this suggests a reduction in the artistic quality of uploads, especially in the nature genre.

\begin{table}[!t]
 \caption{Results: Changes in uploads, Unsplash vs. Instagram} \label{tbl:instagram}
\input{Overleaf/tables/new/instagram}\\
%\noalign{\vskip .5em}
\begin{minipage}{\linewidth}
{\footnotesize
\def\sym#1{\ifmmode^{#1}\else\(^{#1}\)\fi}
\textbf{Note:} 
In column (1), the dependent variable is the number of monthly uploads to Unsplash. In column (2), the dependent variable is the number of monthly uploads to Instagram, and in column (3), the dependent variable is the number of uploads to both platforms. \textit{Treated} indicates whether a user was part of the LITE dataset. Standard errors are clustered at the user level in parentheses. \sym{*} \(p<0.10\), \sym{**} \(p<0.05\) \sym{***} \(p<0.01\) \sym{*}
}
\end{minipage}
\end{table}

The next question is whether the behavioral changes of treated users have broader consequences or are limited to Unsplash. In other words, do photographers who leave Unsplash, reduce their uploads or lower image quality also alter their professional approach or exit the profession entirely, or are their reactions solely a response to perceived unfair treatment by the platform?

To answer this question, we study whether treated users only reduce their contributions to Unsplash or another image-sharing platform as well. As described in section \ref{sec:robust:insta}, for a subset of users, we have access to their upload activities on both Unsplash and Instagram, before and after Unsplash releases data. Panel A (showing upload levels) and Panel C (showing the difference in uploads between Instagram and Unsplash within the treatment group) of Figure \ref{fig:diff_number_photos_insta} suggest that inclusion in the LITE dataset affects only activity on Unsplash, with no spillovers to Instagram. In Table \ref{tbl:instagram}, we provide the corresponding econometric results. First, we see in column (1) that the average treatment effect concerning uploads on Unsplash is about half the size of our baseline (column 1 in Table \ref{tbl:uploads}) in the subsample of users active on Unsplash and Instagram. In column (2), we do not find a significant effect on Instagram uploads and the point estimate is very close to zero. In columns (3) and (4), we combine information from both platforms to compare effect sizes directly. The linear combination in both sets of results, with user fixed effects and with user-platform fixed effects, again indicates that the impact on Instagram is not significant. This implies that users did not switch to Instagram as a new outlet for their photography work, but it does not imply that users stopped publishing their work anywhere online. Overall, our results suggest that adding Unsplash images to an Unsplash-curated AI training dataset affects user behavior on Unsplash, but not elsewhere. More broadly, this leads to the conclusion that the program likely did not incentivize photographers to exit their profession.

\subsection{Implications for AI training data}

\subsubsection{Analysis of variety and novelty measures}

\begin{figure}[!h]
\centering
\caption{Examples of images based on variety and novelty scores} \label{fig:image_examples}
\begin{minipage}
    {.167\linewidth}
    \centering
    \includegraphics[width=\linewidth]{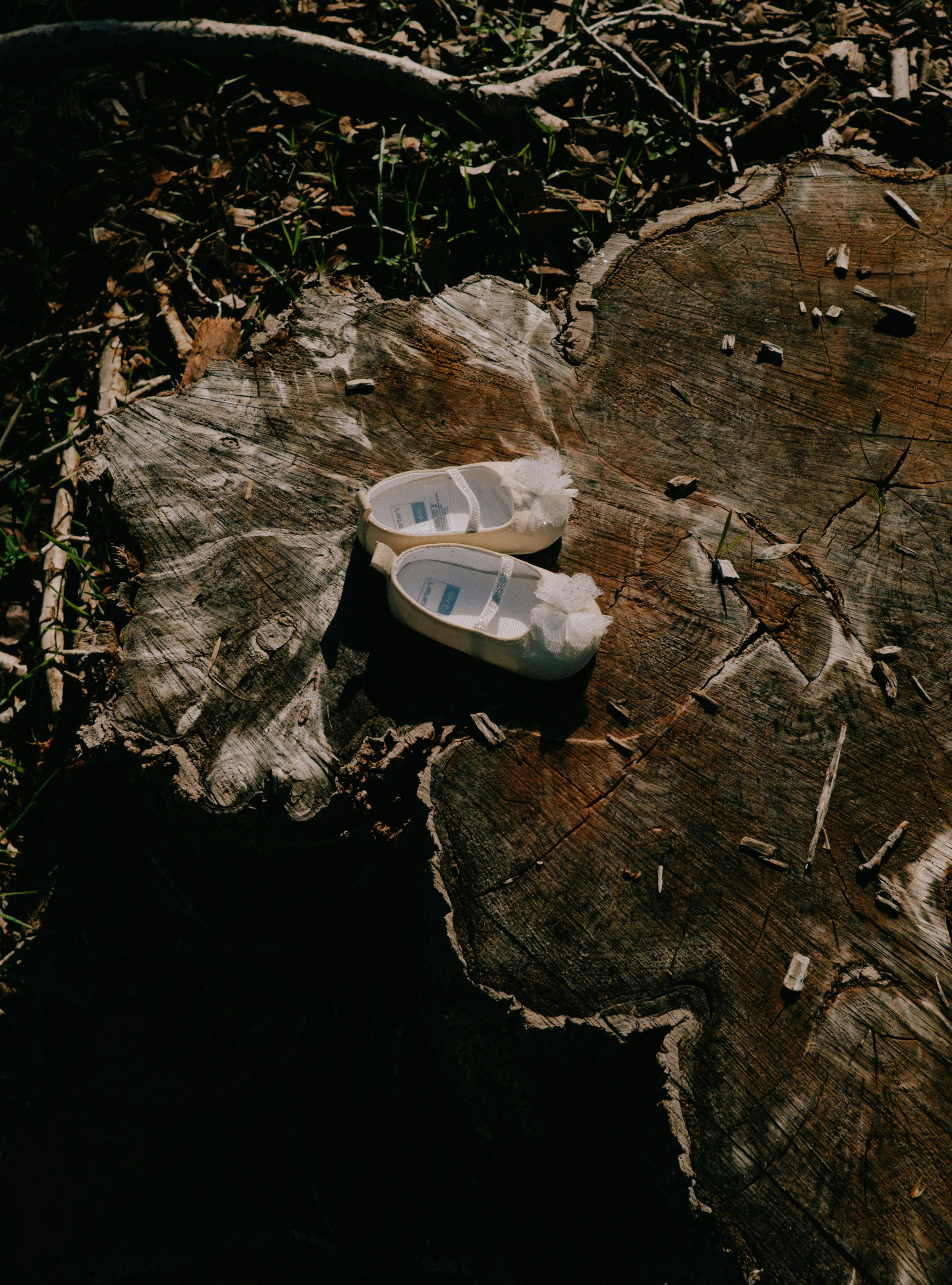}
    \footnotesize{High variety\\
    High novelty}
\end{minipage}
\hfill
\begin{minipage}{.15\linewidth}
    \centering
    \includegraphics[width=\linewidth]{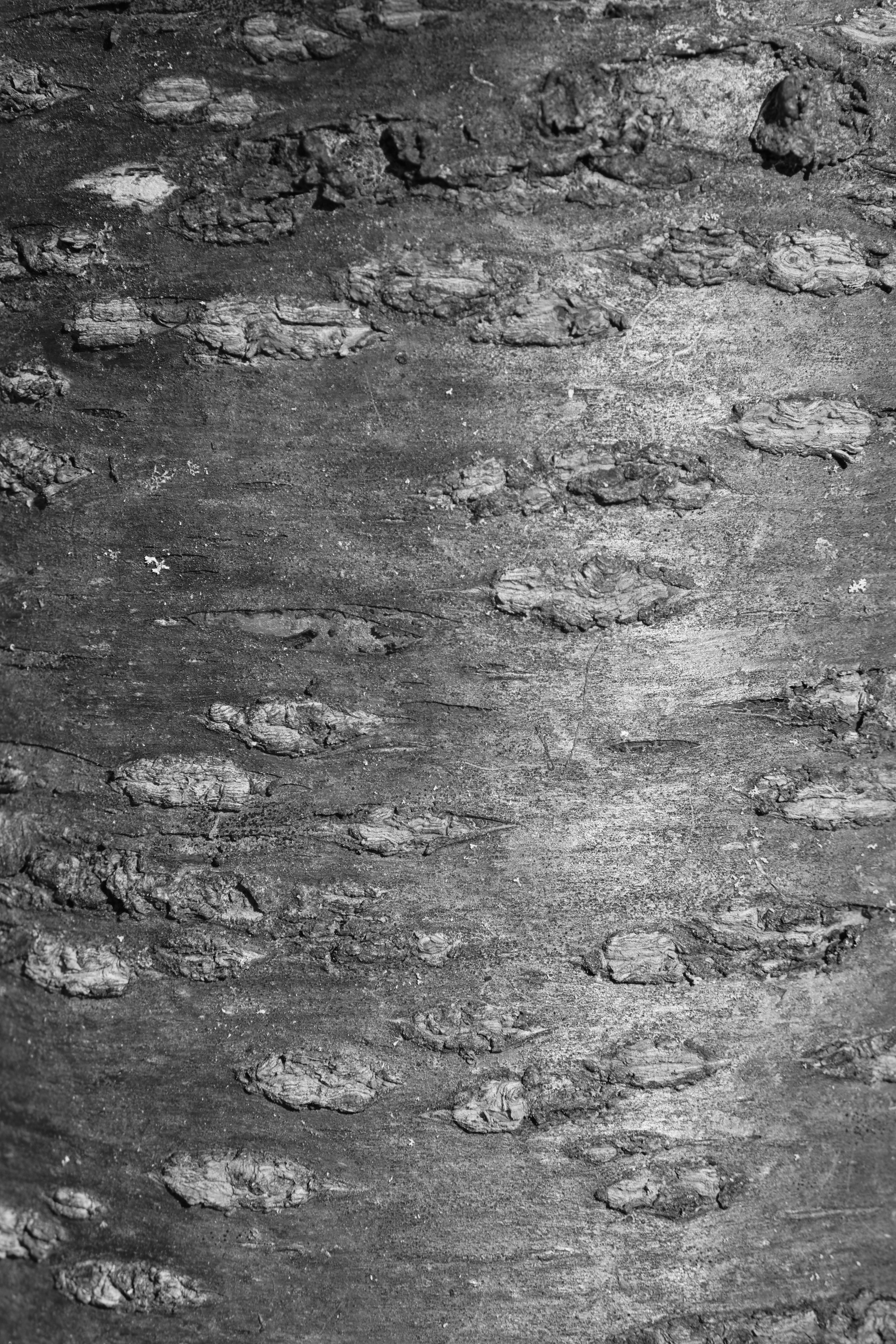}
    \footnotesize{High variety\\
    High novelty}
\end{minipage}
\hfill
\begin{minipage}{.15\linewidth}
    \centering
    \includegraphics[width=\linewidth]{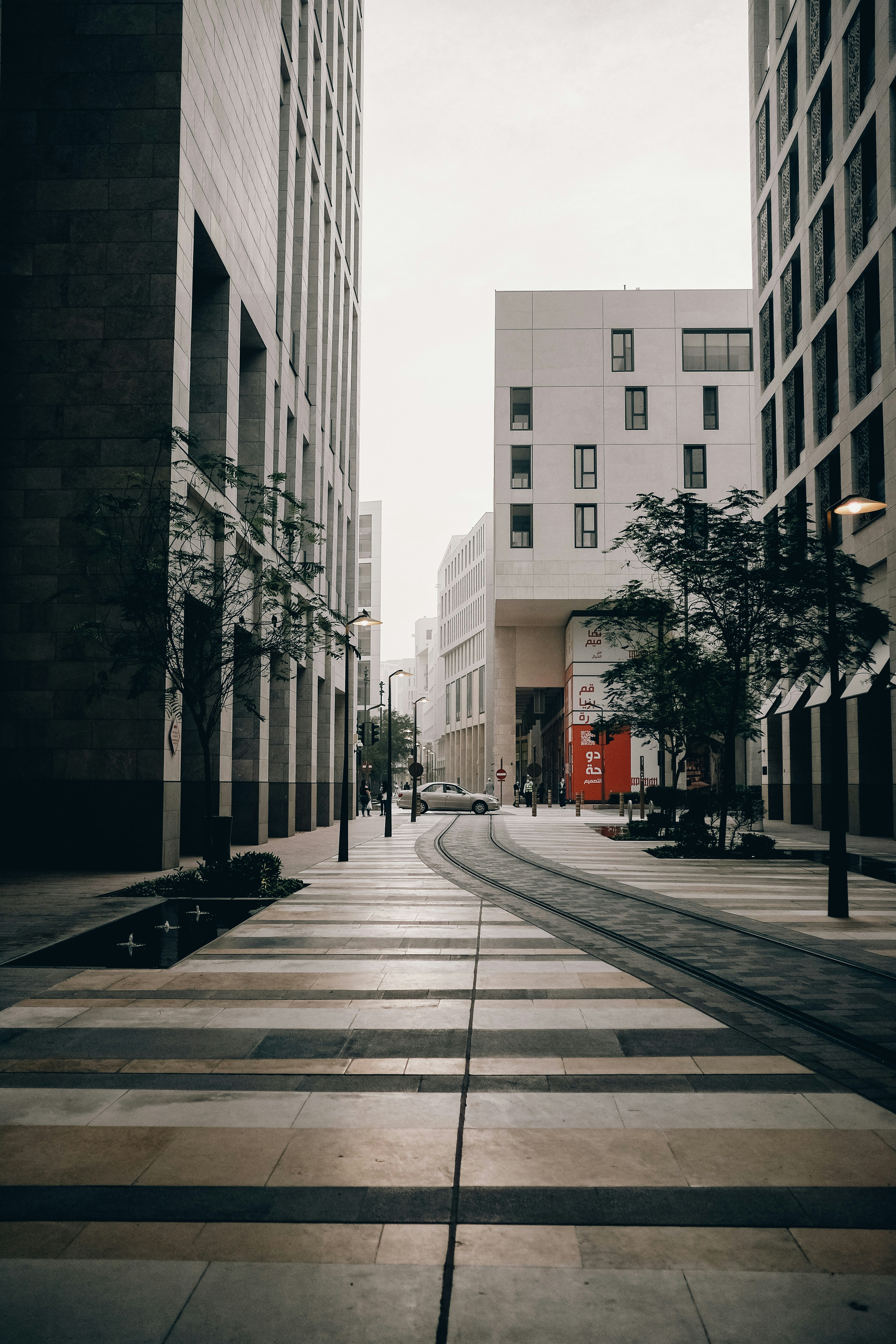}
    \footnotesize{Mean variety\\
    Mean novelty}
\end{minipage}
\hfill
\begin{minipage}{.167\linewidth}
    \centering
    \includegraphics[width=\linewidth]{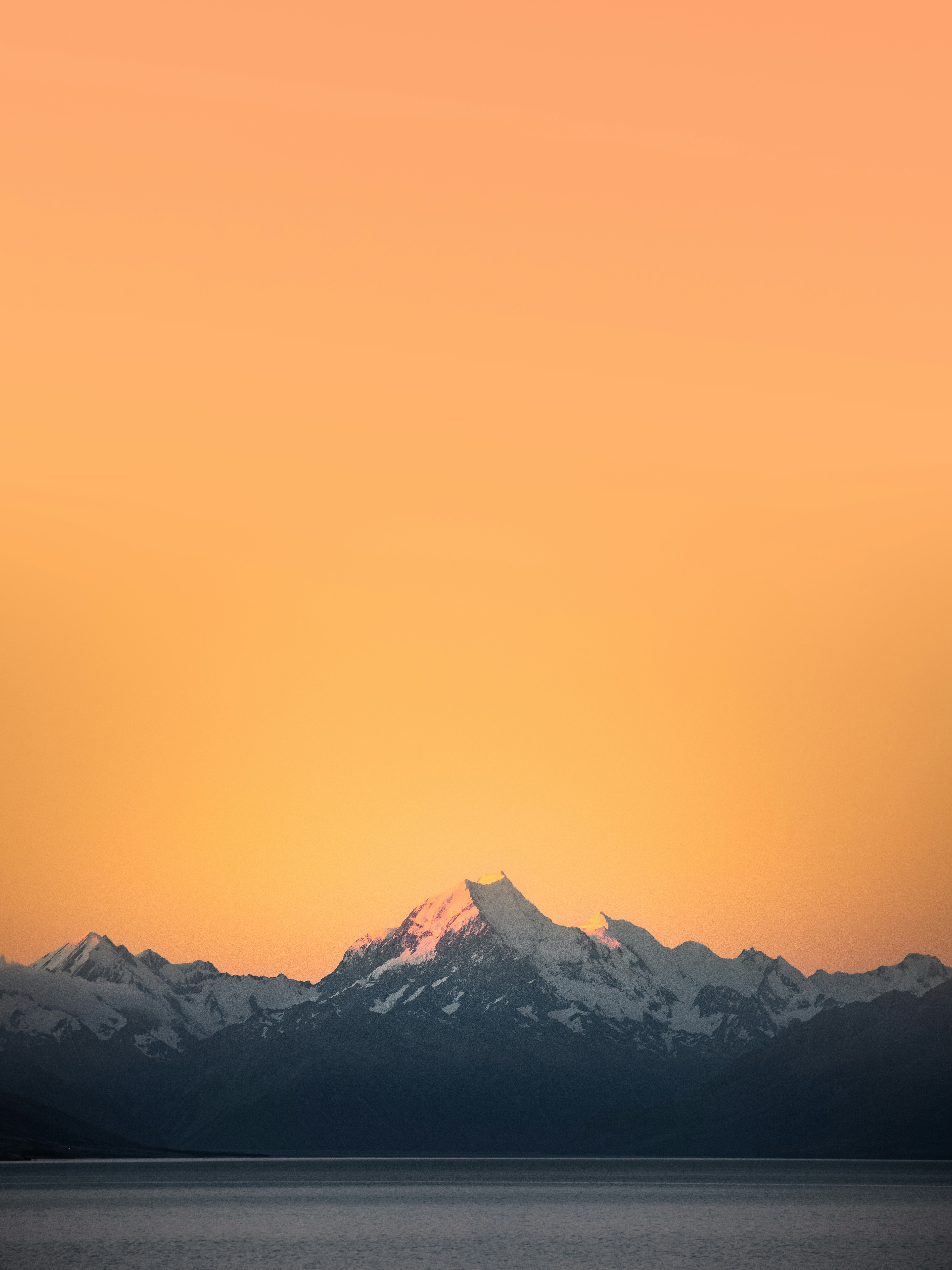}
    \footnotesize{Low variety\\
    Low novelty}
\end{minipage}
\hfill
\begin{minipage}{.15\linewidth}
    \centering
    \includegraphics[width=\linewidth]{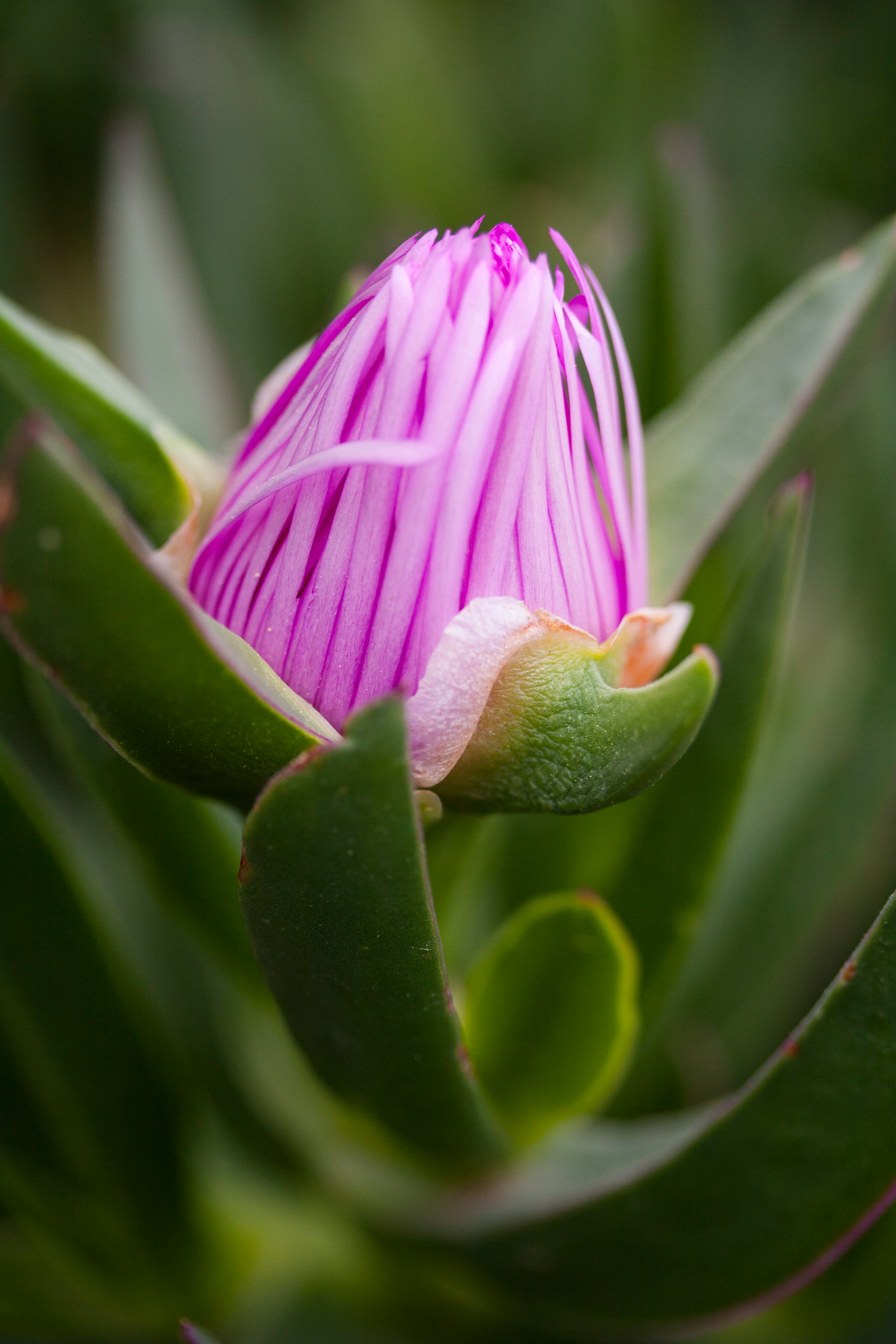}
    \footnotesize{Low variety\\
    Low novelty}
\end{minipage}
\begin{flushleft}
\footnotesize{\textbf{Note:} These examples illustrate how images vary in both variety and novelty according to our four-step scoring procedure. Variety reflects the average similarity across all existing images (high average similarity means low variety score), whereas novelty captures the count of images whose similarity to the new upload exceeds a specified threshold (high count means low novelty score). The first two images (left) exhibit relatively high variety and novelty, the middle image represents mean variety and novelty, and the final two images (right) show low variety and novelty. All examples are drawn from the Unsplash sample as of May 2023. Image credits (from left to right): \href{https://unsplash.com/fr/@marsupialpudding}{Ashe Walker}, \href{https://unsplash.com/@wolfgang_hasselmann}{Wolfgang Hasselmann}, \href{{https://unsplash.com/fr/@reo}}{Emre}, \href{https://unsplash.com/fr/@marekpiwnicki}{Marek Piwnicki}, and \href{https://unsplash.com/fr/@enginakyurt}{Engin Akyurt}.}
\end{flushleft}
\end{figure}

We now turn our attention to the image level.
To investigate whether contributions to Unsplash change after the release of the LITE dataset, we compare the flow of new uploads to the stock of images one year before the launch of the research program. In particular, to capture \textit{variety} as a quality measure of a training dataset, we measure how similar each new upload is relative to all existing images. To capture \textit{novelty}, we count the number of images in the stock of data that are very similar to a focal new upload.

We then compare how variety and novelty evolve over time for treated users and control users, before and after the release of the LITE dataset. We do so based on the keywords that users assign to their uploaded images, which on average gives us 10.28 keywords per image. To assess the similarity between images we proceed in four steps of NLP. 

In step one, we aggregate keywords for each image and train a Word2Vec model on these aggregated keywords to generate a vector representation for each unique keyword. Herein, the model is configured to create 100-dimensional vectors for each keyword, with a context window of 5 words and a minimum word frequency threshold of 1, ensuring that even rare keywords are included in the model.
In step two, we apply the Word2Vec model to each image, converting the keywords into vectors, and then calculate the average vector to represent the image's semantic content.
In step three, we calculate cosine similarity scores between the vector representation of a particular image and the vector representations of all existing images.
Finally, in step four, we compute the average similarity across all images and count the number of images that exceed a specified similarity threshold, respectively.

We perform steps 1 and 2 for the stock of images uploaded by users that will be treated eventually as well as for the stock of images uploaded by users in the control group. We define the stock as all images uploaded before June 25, 2019.
We perform steps 1-4 for all images uploaded by treated and control users after June 25, 2019. This allows us to compare new uploads starting one year before the start of Unsplash's research program to new uploads after the release of the first LITE dataset, across users in the treatment and in the control group. In step 4, \textit{variety} is defined as the average cosine similarity between a new upload and the images in the stock, while \textit{novelty} is defined as the count of stock images whose similarity to the new upload exceeds a threshold (here 0.8 or 0.9). Formally, if $s(\mathbf{x}_i,\mathbf{x}_j)$ is the cosine similarity between the embeddings of images $i$ and $j$, and $\theta$ is a chosen threshold, then the novelty measure for image $i$ is:
\[
\sum_{j \in \text{Stock}} \mathbb{I}\bigl[s(\mathbf{x}_i,\mathbf{x}_j) > \theta\bigr],
\]
where $\mathbb{I}[\cdot]$ is an indicator function. A higher count indicates that $i$ has more ``near neighbors'' in the stock and is thus \textit{less} novel. Conversely, a smaller count implies \textit{greater} novelty. Figure \ref{fig:image_examples} presents five images illustrating different segments of the variety--novelty space. From left to right, the first two show images with high variety and high novelty scores, the middle image represents mean variety and novelty scores, and the final two exhibit low variety and low novelty scores. This highlights how images can vary in their average similarity to the stock (variety) and the number of ``near neighbors'' above a similarity threshold (novelty).

\begin{figure}[!t]
\caption{Changes in uploads, similarity of images} \label{fig:sim_levels_differences}
\begin{minipage}{.5\linewidth}
\footnotesize \textit{A: Variety - Levels of average similarity}\\
        \includegraphics[trim=10 0 0 0 , clip, width=\textwidth]{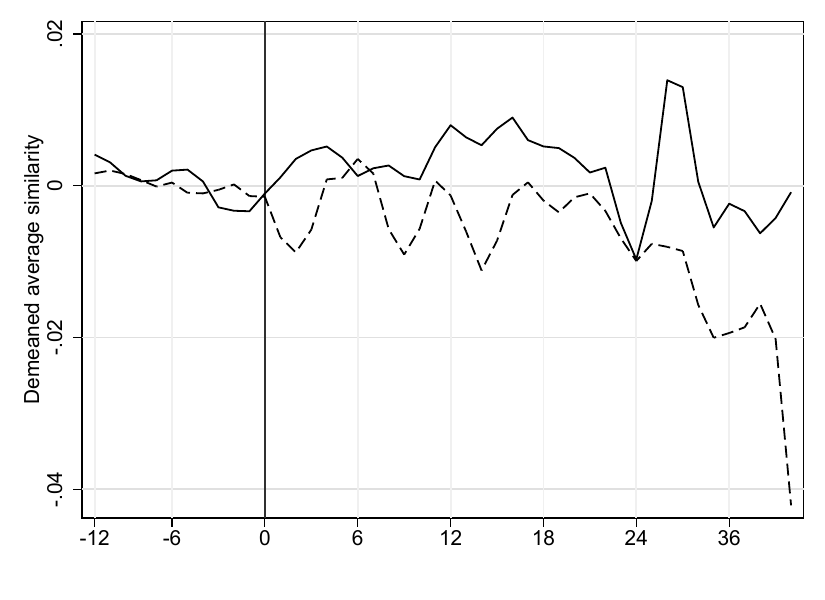}\\
\end{minipage}
\begin{minipage}{.5\linewidth}
\footnotesize \textit{B: Variety - Differences in average similarity}\\
        \includegraphics[trim=10 0 0 0 , clip, width=\textwidth]{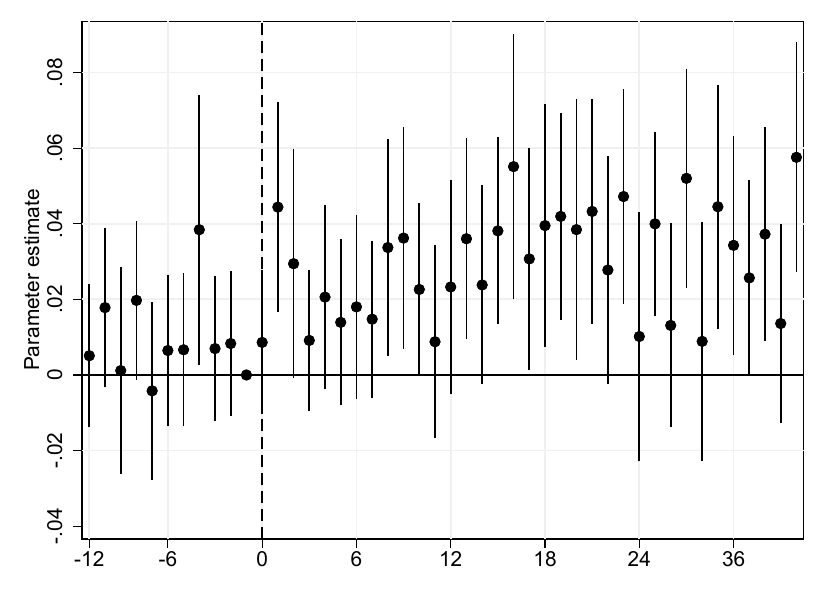}\\
\end{minipage}\\
\begin{minipage}{.5\linewidth}
\footnotesize \textit{C: Novelty - Levels of number of very similar images}\\
        \includegraphics[trim=10 0 0 0 , clip, width=\textwidth]{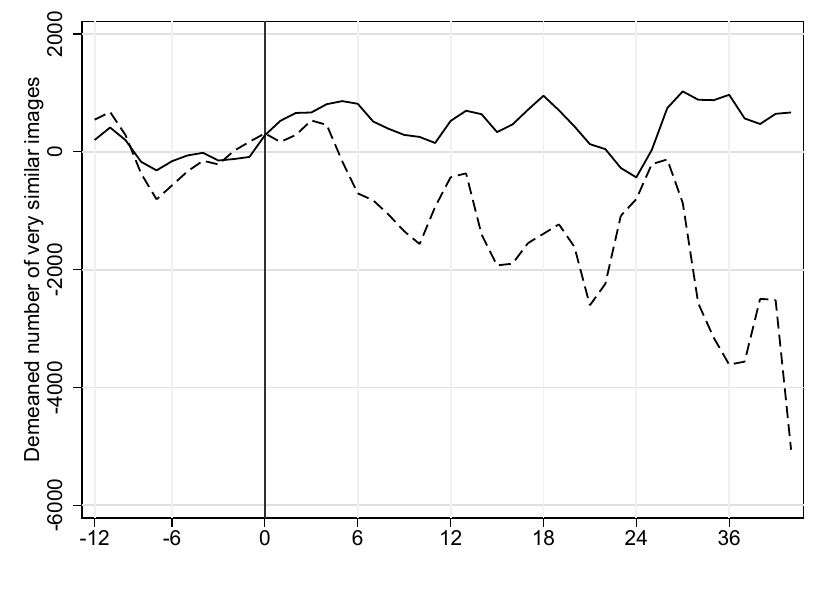}\\
\end{minipage}
\begin{minipage}{.5\linewidth}
\footnotesize \textit{D: Novelty -  Differences in number of very similar images}\\
        \includegraphics[trim=10 0 0 0 , clip, width=\textwidth]{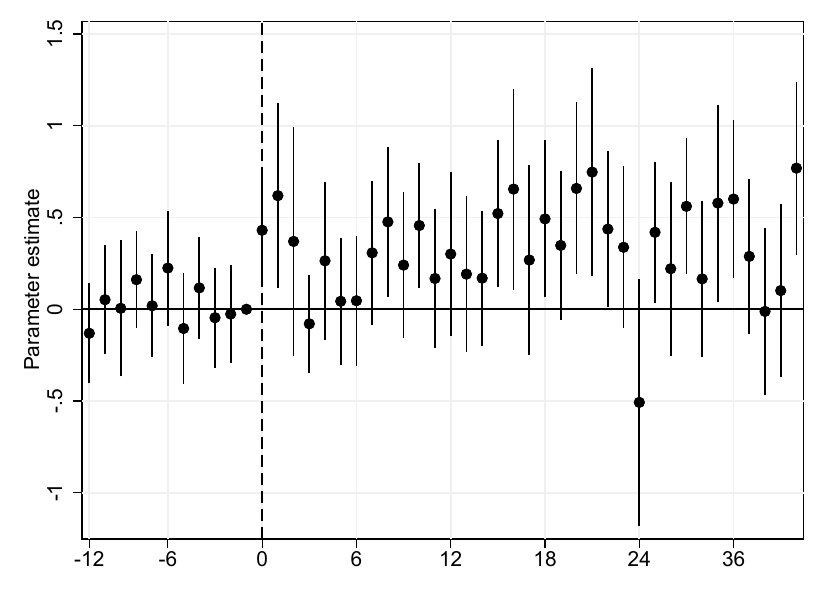}\\
\end{minipage}\\
{\footnotesize
\textbf{Note:} In Panel A, we depict the demeaned average similarity of uploaded images by users in the treatment group (solid) and by users in the control group (dashed). In Panel C, we plot the demeaned number of images uploaded by treated users (solid) and control users (dashed) that are very similar to the existing stock (threshold 0.8). We use the group-specific pre-period mean number of uploaded images to demean (0.38 and 8,403 for the treatment group, and 0.39 and 8,566 for the control group). In Panels B and D, we plot OLS estimates of the $\delta_{\tau}$ coefficients obtained from estimating $Y_{it} = \sum_{\tau \in T} \nu_{\tau} + \sum_{\tau \in T} \delta_{\tau} \left( \gamma_{\tau} \times  Treated_{i} \right)  + \varepsilon_{it}$, with the number of uploads of user $i$ per month $t$ as the dependent variable. The dots reflect month-specific point estimates comparing treatment users to control users. Standard errors are clustered at the user level, and bars indicate 90\% confidence bands.
In all panels, the vertical line indicates the creation of LITE 1.0.0, and numbers on the horizontal axis indicate the number of months before and after that.
}
\end{figure}

\begin{table}[!t]
 \caption{Results: Changes in uploads, similarity of images} \label{tbl:uploads_image_sim}
\input{Overleaf/tables/new/uploads_image_sim}\\
%\noalign{\vskip .5em}
\begin{minipage}{\linewidth}
{\footnotesize
\def\sym#1{\ifmmode^{#1}\else\(^{#1}\)\fi}
\textbf{Note:} 
The dependent variable in columns (1) and (2) is computed as the average similarity between an uploaded image and the stock of images uploaded by treated vs. control users up until one year before the creation of the LITE dataset. In columns (3)--(6), the dependent variable is the log+1 count of images from the stock of images uploaded by treated/control users up until one year before the creation of the LITE dataset that have a cosine similarity score of larger than 0.8 or 0.9 relative to an uploaded image. \textit{Treated} indicates whether a user was part of the LITE dataset. Fixed effects for upload month of the image and user account in columns (2), (4), and (6). Standard errors in parentheses, White-robust in columns (1), (3), and (5), and clustered at the user level in columns (2), (4), and (6). \sym{*} \(p<0.10\), \sym{**} \(p<0.05\) \sym{***} \(p<0.01\)
}
\end{minipage}
\end{table}

Panel A of Figure \ref{fig:sim_levels_differences} shows that the demeaned average similarity to the pre-existing stock of images decreases less for images uploaded by treated users than for images uploaded by control users. In other words, treated users upload works that vary less and are more similar to the existing stock of images. This is confirmed more formally in Panel B, where we plot estimates of the difference between the treatment and control group in each month, before and after the release of the LITE dataset, from a model without user-fixed effects. In Panels C and D, we turn to our novelty measure, a count of the number of very similar images. Again, we see that the decrease in very similar images uploaded by users in the treatment group is less steep than in the control group. Finally, the lack of statistically significant deviations from 0 in the pre-period (all month-specific differences are jointly not significantly different from zero) supports the parallel trends assumption of the difference-in-differences model. 

We quantify these differences in Table \ref{tbl:uploads_image_sim}.
We find that it is not users who change the type of images they upload, but the difference in the type of uploaded image derives from a change in the composition of users. In the specifications with user-fixed effects in columns (2), (4) and (6), we do not find much evidence that the release of the LITE dataset caused new uploads to differ from existing uploads. This is especially the case concerning the number of highly similar images in the existing stock. Hence, users do not change the types of images they upload. However, looking at the results from specifications without user-fixed effects in columns (1), (3), and (5), we do see that uploaded images become more similar to the existing stock. Column (1) suggests an increase of about 5\% in average similarity for uploads by treated users, whereas columns (3) and (5) suggest an increase of about 29\% and 14\%, respectively, in the number of highly similar images.

\subsubsection{Link to AI output quality}

Having quantified how user behavior changes the quantity, variety, and novelty of the flow of data, we now discuss the potential knock-on effects on AI output quality. First, ample evidence suggests that algorithmic performance is positively related to dataset size \citep[see e.g.,][]{budach2022effects, valavi2022time2, Peukert2024Editor}. This corresponds to evidence showing that the AI industry is vying to secure access to big datasets \citep{beraja2023data}. Although returns to scale are decreasing, a rough linear approximation of the estimates suggests that a 2\% reduction in sample size, like the one we find in our context, implies a 1.4--2\% reduction in algorithmic performance (see Figure 2 in \citealp{Peukert2024Editor} and Figure 12 in \citealp{budach2022effects}).

Second, regarding the quality of data, a growing but relatively small area of research suggests that poor or degrading data quality decreases the performance of AI models (see \citealp{zha2023data} for a recent survey). Experimental studies show that dynamic changes in the distribution of training data can degrade algorithmic performance (related concepts in Machine Learning are \textit{concept drift}, \textit{recursion}, and \textit{model collapse}). For example, in an exercise comparable to our empirical setting, \cite{cai2021online} and \cite{jain2023learning} study the computer vision problem of geolocalization. In essence, they test the prediction accuracy of a model that learns the geographic location depicted in photos, using both image data as well as geolocation data from sensors in the camera device. Resembling the example of a changing city skyline discussed above, the authors show that the prediction accuracy of a model trained on a batch of data from a particular time period is highest when images in the test dataset are from the same or previous ranges, but accuracy quickly declines when images in the test dataset are from subsequent periods. The natural shift in the distribution of images implies a reduction in model accuracy of about 4--9\% over three years (see Figure 1 in \citealp{jain2023learning}). Based on this, we can speculate that a reduction in novelty of about 13\% to 28\% beyond the natural rate, as we have estimated above, would translate into a decrease in output quality of 4.5\% to 11.5\%.

A potential solution is to impute missing or low-quality data with generated training data, i.e. synthetic data, to solve the issues of data quantity and quality \citep{lucini2021real}. Recent papers in law (e.g., \citealp{lee2024}) and computer science (e.g. \citealp{gerstgrasser2024model}) discuss the potential for and address the concerns related to synthetic data and AI performance. However, a decrease in the variety of user-generated data, as in our empirical context, could still be detrimental to AI model performance. 

A recent study points out how the injection of synthetic data into training data may not degrade AI outputs as long as it adds to, rather than replaces, the stock of data (see e.g., \citealp{gerstgrasser2024model}). However, assuming that the value of past data can decay over time, with rates depending on the use case or topic of interest \citep{valavi2022time}, adding synthetic data effectively does replace the stock of training data over time. Thus, if we interpret AI-generated, synthetic, content as information similar to the input data, we can use the empirical estimates in this literature to speculate how a reduction in the variety in the flow of human-generated data translates into algorithmic performance. Experimental evidence suggests that adding AI-generated data to the training data implies a reduction in ``accuracy'' of 5\% to 22\% (see Figure 3 in \citealp{gerstgrasser2024model}, when Cross-Entropy Loss is interpreted as a measure of accuracy for the sake of a simple argument). Assuming that AI-generated content is identical to input data, our estimate of a 5\% increase in similarity would reduce accuracy by 0.25\% to 1.4\%.

In conclusion, proposals that point out the potential to train models with synthetic data to circumvent human input and thus copyright-related challenges (e.g., \citealp{lee2024}) need to be interpreted with caution. In situations where the value of the stock of data decays over time, issues related to AI model performance and eventual collapse can't be addressed by synthetic data \citep{ valavi2022time, cai2021online, jain2023learning, shumailov2024ai}. While the temporal dynamics of data are considered exogenous in computer science (concept drift), the social sciences recognize it as endogenous \citep{Quinn2025}, subject to the behavior of and incentives for the humans that produce the information. Hence, policy solutions need to be guided by the insight -- i.e. strategic individual and organizational behavior govern the supply, and therefore quality of training data and eventually AI output.

\section{Discussion}

Creators across various industries have exhibited comparable responses to AI models trained on their work, reinforcing the broader applicability of our findings.
Authors have expressed outrage upon discovering that AI models were trained on their copyrighted books without consent or compensation. For instance, Australian publisher Black Inc asked its authors to sign agreements allowing their works to be used for AI training, leading to significant backlash from the literary community.\footnote{See \url{https://www.theguardian.com/books/2025/mar/05/black-inc-melbourne-publisher-ai-agreements-writers-anger}.}
Musicians have also raised concerns about AI-generated songs potentially diluting their revenue streams and reducing incentives for new compositions.\footnote{See \url{https://www.billboard.com/pro/labels-artists-sign-petition-opposing-ai-training-creative-works/}.} Over 1,000 artists released a silent album titled ``Is This What We Want?'' in protest against UK government proposals to allow AI companies to use copyright-protected works without permission.\footnote{See \url{https://www.theguardian.com/technology/2025/feb/25/kate-bush-damon-albarn-1000-artists-silent-ai-protest-album-copyright}.}
A controversy on Instagram unfolded when artists learned their posts might be used to train Meta's AI models, causing some users to migrate to platforms promising not to engage in AI data mining. In response, Meta has faced pressure to clarify or revise its policies regarding AI's use of user content.\footnote{See \url{https://www.washingtonpost.com/technology/2024/06/06/instagram-meta-ai-training-cara/}.}

In what follows, we discuss key implications for managers and policy that arise from our results .

\subsection{Managerial implications} \label{section:discussion:management}

Realizing the value of their content for genAI training purposes, platforms are increasingly partnering with AI firms to license their content archives to AI companies.\footnote{OpenAI signed several agreements with newspapers such as the Guardian or Newscorp. See \url{https://openai.com/index/openai-and-guardian-media-group-launch-content-partnership/} and \url{https://www.nytimes.com/2024/05/22/business/media/openai-news-corp-content-deal.html}.} However, decision-makers should be mindful of data-sharing terms, as misaligned or exploitative policies can drive contributors -- and their content -- elsewhere. Our findings show that user responses to the exploitation of their content in AI training meaningfully affects both the quantity and quality of contributions. From a managerial perspective, this raises questions about how platform operators and other industry stakeholders can capitalize on AI's potential without sacrificing the supply of human-generated content.

One approach is to provide multiple levels of data licensing and usage rights that incentivize continued engagement. For example, in October 2022 Unsplash introduced Unsplash+, a paid subscription service that explicitly forbids using images for machine learning, AI, and biometric tracking technology, and offers contributors monetary compensation for permitted uses.\footnote{See \url{https://unsplash.com/plus/license} and \url{https://unsplash.com/blog/contribute-to-unsplash/}.} Similarly, Shutterstock, another stock photography company, announced a ``contributor fund'' to compensate artists whose work is integrated into generative AI datasets.\footnote{See \url{https://support.submit.shutterstock.com/s/article/Shutterstock-ai-and-Computer-Vision-Contributor-FAQ?language=en_US}.} Alternatively, platforms can implement ``opt-out'' tags for AI crawlers or introduce specialized ``AI-ready'' tiers, which give creators enhanced levels of control. In a similar vein, AI companies are acknowledging creators' concerns. After being confronted with artist complaints and a series of impending lawsuits, OpenAI developed tools to give content creators more control, such as allowing artists to opt out of having their art used in future model training.\footnote{See \url{https://www.wired.com/story/openai-olive-branch-artists-ai-algorithms/}.} As of this writing, it remains unclear whether these attempts will prove fruitful. Hence, we need additional research into how platforms can profit from genAI's appetite for data without deterring contributor participation.

\subsection{Market-level implications}

Our analysis shows that adding user-generated content in AI training datasets can lead to behavior that negatively impacts the quantity, variety, and novelty of contributions, thereby reducing AI output quality. If policymakers and industry stakeholders do not adequately address creators' concerns -- particularly around fair compensation and rights management -- both rights holders and AI-driven innovation may suffer in the long run. One potential path forward could be an automated dynamic compensation model that pays royalties based on the usage and commercial success of AI models trained with human-created works.
On the one hand, a usage-based remuneration system would compensate contributors whenever their works are used in the training phase of an AI model (see \citealp{Wang2024Economic} for a technical proposal), much like music royalties that are paid out for each song played. Further, our results show that more intensely treated users display a more pronounced reaction. This highlights that a one-size-fits-all approach to licensing human-generated work for AI training datasets can lead to unintended heterogeneous effects on the flow of training data. Consequently, a licensing scheme that ensures a continued flow of data may need to compensate more active contributors to a larger degree.
On the other hand, outcome-based bonuses could reward contributors if AI models built on their data achieve commercial success or lead to significant technological breakthroughs -- akin to receiving royalties when a song is reused. Such a dual compensation mechanism would ensure that contributors are rewarded not just for the use of their data but also for the value it generates. Deploying such a system at scale would require advanced tracking technologies (e.g., digital fingerprinting) from training to output generation, as well as trusted intermediaries to manage potentially complex and large-scale transactions across various stakeholders in the AI ecosystem.
%To implement such a model at scale, advanced tracking technologies are required that can monitor the usage of data throughout an AI's life cycle -- from training to output generation. Here, digital fingerprinting can help in uniquely identifying data contributions across different AI models and their outputs, allowing for accurate compensation calculations. Trusted intermediaries would ensure that every instance of data usage is recorded securely. The administration of compensation models can be complex, involving numerous transactions and the need for widespread coordination across various stakeholders in the AI ecosystem. 

The existing data economy around the online advertising industry can serve as inspiration. Data auctions and data exchange platforms may provide efficient solutions. Data auctions facilitate efficient price discovery as AI developers and firms bid for the rights to use datasets, reflecting real-time market supply and demand. This competitive bidding process incentivizes creators to produce diverse, high-quality data likely to attract higher bids. The transparent nature of auctions provides clear benchmarks for data valuation, enhancing overall market transparency. Simultaneously, data exchange platforms could serve as centralized markets similar to stock exchanges but dedicated to trading data rights. Here, data creators can list their datasets, allowing buyers to purchase rights for specific uses. 
Much like rights collection societies in music, authorship, and the visual arts, these platforms would standardize transactions, reduce the complexity of individual agreements, and help ensure regulatory compliance. This may increase the liquidity of data as a commodity, making it accessible for AI developers when needed and expanding the market reach for data creators. By making pricing, supply, and demand visible, such platforms could enhance market efficiency, allowing buyers and sellers to make informed decisions based on up-to-date market information. 

For policymakers, finding the right solution to the ongoing discussions surrounding AI training data and copyright will be crucial. Failure to do so could potentially lead to adverse outcomes, such as AI firms relocating their model training operations to copyright-friendly jurisdictions like Israel or Japan \citep{fiil2022legal}. %\footnote{See \url{https://www.economist.com/business/2024/04/14/generative-ai-is-a-marvel-is-it-also-built-on-theft} and \url{https://www.economist.com/schools-brief/2024/07/23/ai-firms-will-soon-exhaust-most-of-the-internets-data}.}
Given the global competition in the regulation of digital markets, countries compete not only through tax incentives but also by offering attractive legal frameworks for AI research and development.

%%%%%%%%%%%%%%%%%%%%%%%%
\section{Conclusion}
%%%%%%%%%%%%%%%%%%%%%%%% 

We investigate the behavioral reactions of creators when their works are made available for the training of commercial AI applications and how this affects the continued supply of content. Our empirical context provides us with a natural experiment on the photography platform Unsplash we exploit to derive empirical insights. Our results imply that the release of an AI training dataset for commercial use, covering roughly 5\% of the platform's contributors, reduced overall uploads (and, therefore, the flow of AI training data) by 2\%. It also decreased the variety (deviation from the average image in the stock) of the average upload by 5\%, and the novelty of the average upload (repetition of very similar images in the stock) by 14-28\%.
 
Our findings echo prior work documenting adverse reactions from content creators when AI is introduced into their domains \citep{huang.2023.GenerativeAIContentCreator,lin2024,Quinn2025} and correspond to the broader reactions of rights holders, including writers, musicians, and visual artists.

Prior literature has focused on the demand-side implications of AI \citep{Felten2023,yiu2024ai,Brynjolfsson2025,Demirci2025}, whereas studying the upstream supply of data forms the contribution of our research. We present empirical evidence for theoretical arguments on AI-induced market changes \citep{Wang2024Economic,yang2024generative,gans2025copyright}, and provide the microfoundations -- causal evidence on the behavioral responses of data contributors -- for the broader trend in the decreasing availability of open data on the web \citep{Longpre2023}. Our research complements previous empirical studies on how changes in the supply of human-generated data may impact the quality of AI training datasets \citep{Jia2025}. Understanding these dynamics is crucial as AI development is moving forward in an environment where high-quality, diverse, and legally accessible data may become increasingly scarce.

The computer science literature suggests that synthetic data -- AI-generated rather than human-created -- could reduce the reliance on human-generated content \citep{ho2024algorithmic}. However, studies warn that excessive use of synthetic data in training can degrade model performance, leading to what is known as ``model collapse'' \citep{valavi2022time2,shumailov2024ai}. Hence, technical solutions seem difficult, which implies that setting the right incentives for human creators to continue contributing to AI training datasets forms a first-order problem for managers and policymakers. The solutions need to include compensation mechanisms and data usage frameworks that align the incentives of creators, content platforms, AI developers, and the users of AI models. Concerning the aspects mentioned above, the current policy appears misguided. For example, the existing regulatory framework in Europe requires high-quality training datasets that are compliant with copyright law, which allows rights holders to opt out of certain uses of their works.\footnote{See EU AI Act, Article 10(3), \url{https://eur-lex.europa.eu/legal-content/EN/TXT/?uri=OJ:L_202401689\#art_10} and EU Copyright Directive, Article 4, \url{https://eur-lex.europa.eu/legal-content/EN/TXT/?uri=CELEX:32019L0790\#art_4}.} Our findings highlight that such a regime can introduce bias in training datasets and lower the quality of AI model outputs.

%Limitations and future research
Our paper provides a first step to understanding the supply side of AI training data, but further research will be required moving forward. From a technical standpoint, we need direct evidence on how a decrease in the flow of data -- specifically in its quantity, variety, and novelty -- affects AI model performance. From a practical standpoint,  additionnal empirical evidence on how the compensation of creators influences the quantity and quality of training data to design effective compensation mechanisms is of high importance.

%potential further limitations: 
%% 1) Can't assess motives of contributors 2) not directly relevant to the points we are trying to communicate - more about breaking ground on potential negative consequences of making user-generated content available for AI training 3) Would benefit policy recommendations if future research dives into motives and whether compensating contributors would be able to repair decrease in activity
%% 1) we assume a linear relationship in ATT as size of treated group increases 2) We are likely estimating a conservative lower bound though, given herding behavior in platforms and communities (Oh and Jeon, 2007) 3) Future research could investigate network dynamics

% \textbf{Author contributions}:

% % FA: Background Research, Data Collection
% % JH: Background Research, Data Analysis, Writing
% % FK: Background Research, Data Collection, Writing 
% % CP: Idea, Background Research, Data Collection, Data Analysis, Writing, Funding Acquisition
% % AS: Background Research, Data Analysis, Writing

\newpage
\singlespacing
\bibliographystyle{itaxpf}
\bibliography{references}

\newpage
\clearpage
\doublespacing
\section*{Appendix}

\setcounter{table}{0}
\setcounter{figure}{0}
\renewcommand{\thetable}{A.\arabic{table}}
\renewcommand{\thefigure}{A.\arabic{figure}}
\setcounter{lstlisting}{0}
\renewcommand{\thelstlisting}{A.\arabic{lstlisting}}

\singlespacing

\captionsetup{type=lstlisting} % Declare that caption refers to a listing
\caption{Python script to download all images in LITE} \label{lst:image_download}
\begin{lstlisting}[language=python, frame=single, framerule=1pt]
import os
import time
import requests
import pandas as pd
from tqdm import tqdm
from concurrent.futures import ThreadPoolExecutor, as_completed

file_path = "unsplash-research-dataset-lite-1.0.1/photos.tsv000"
df = pd.read_csv(file_path, sep="\t")
output_dir = "unsplash-research-dataset-lite-1.0.1/images"
os.makedirs(output_dir, exist_ok=True)

# Function to download a single image
def download_image(row):
    image_url = row.get("photo_image_url")
    photo_id = row.get("photo_id")
 
    if pd.notna(image_url) and pd.notna(photo_id):
        try:
            response = requests.get(image_url, stream=True, timeout=10)
            if response.status_code == 200:
                image_path = os.path.join(output_dir, f"{photo_id}.jpg")
                with open(image_path, "wb") as f:
                    for chunk in response.iter_content(1024):
                        f.write(chunk)
                return True
            except Exception:
                pass
        return False
 
max_threads = 10
print("Starting image downloads in parallel...")

start_time = time.time()

with ThreadPoolExecutor(max_workers=max_threads) as executor:
    futures = [executor.submit(download_image, row) for _, row in df.iterrows()]
    for _ in tqdm(as_completed(futures), total=len(futures)):
        pass
 
elapsed_time = time.time() - start_time
print(f"Downloaded {len(df)} images in {elapsed_time:.2f} seconds."
\end{lstlisting}\vspace*{-1em}
{\footnotesize \singlespacing
\textbf{Note:} Python script to download all images released in the LITE version 1 dataset. With this script below, the authors were able to download 24,989 images of the first version of the LITE dataset (81.04GB) in 21 minutes on a standard MacBook Pro (11 images were no longer
available).
}

\clearpage
\captionsetup{type=lstlisting} % Declare that caption refers to a listing
\caption{Database query selecting images for LITE} \label{lst:query}
\begin{lstlisting}[language=SQL, frame=single, framerule=1pt]
SELECT DISTINCT photos.id FROM photos
JOIN tags ON tags.photo_id = photos.id
WHERE tags.keyword = 'nature' -- The image is tagged with "nature" 
      AND tags.confidence > 0.9 -- We are more than 90% confident that "nature" is present in that photo
      AND photos.curated =  TRUE -- image has been curated by the editorial team
      AND photos.status = 1 -- image is still available on the platform
LIMIT 25000
\end{lstlisting}\vspace*{-1em}
{\footnotesize \singlespacing
\textbf{Note:} This query was sent to the authors by the head of data science at Unsplash.
\vspace*{1em}
}

\begin{table}[!h]
 \caption{Results: Likelihood of remaining on the platform (interacted)} \label{tbl:survival_image_usertypes}
\input{Overleaf/tables/new/survival_image_usertypes}\\
%\noalign{\vskip .5em}
\begin{minipage}{\linewidth}
{\footnotesize
\def\sym#1{\ifmmode^{#1}\else\(^{#1}\)\fi}
\textbf{Note:} We estimate a linear probability model. The dependent variable is an indicator for whether an image remains on Unsplash as of May 2023. The sample in column (3) is smaller because data on ``ForHire'' is not available for all users. The variable \emph{Treated} indicates whether a user was part of the LITE dataset. We compare survival rates over 2.5 years (from Aug 2020 to May 2023).  We include fixed effects for the image's age (in months) in all specifications. Standard errors clustered at the user level in parentheses. \sym{*} \(p<0.10\), \sym{**} \(p<0.05\) \sym{***} \(p<0.01\)
}
\end{minipage}
\end{table}

% \begin{table}[!h]
%  \caption{Results: Changes in uploads (Time trends)} \label{tbl:uploads_trends}
% \input{Overleaf/tables/new/uploads_trends}\\
% %\noalign{\vskip .5em}
% \begin{minipage}{\linewidth}
% {\footnotesize
% \def\sym#1{\ifmmode^{#1}\else\(^{#1}\)\fi}
% \textbf{Note:} 
% The dependent variable in all columns is the total number of uploads per month. \textit{Treated} indicates whether a user was part of the LITE dataset. 
% Column (1) replicates the preferred baseline specification (column 2 of Table \ref{tbl:uploads}). Column (2) only includes fixed effects for the user account. Column (3) additionally includes time-fixed effects at the yearly level. Column (4) additionally includes group-specific time trends (at the monthly level). Standard errors clustered at the user level in parentheses. \sym{*} \(p<0.10\), \sym{**} \(p<0.05\) \sym{***} \(p<0.01\)
% }
% \end{minipage}
% \end{table}

\begin{table}[!h]
 \caption{Results: Changes in uploads (Weighted)} \label{tbl:uploads_weighted}
\input{Overleaf/tables/new/uploads_weighted}\\
%\noalign{\vskip .5em}
\begin{minipage}{\linewidth}
{\footnotesize
\def\sym#1{\ifmmode^{#1}\else\(^{#1}\)\fi}
\textbf{Note:} 
The dependent variable in column (1) is the total number of uploads per month, in column (2) it is an indicator of whether a user account has uploaded at least one image in a given month, and it is the log number of uploads in column (3). \textit{Treated} indicates whether a user was part of the LITE dataset. Fixed effects for upload month of the image and user account. Standard errors clustered at the user level in parentheses. \sym{*} \(p<0.10\), \sym{**} \(p<0.05\) \sym{***} \(p<0.01\)
}
\end{minipage}
\end{table}

\begin{table}[!h]
 \caption{Results: Changes in uploads, alternative control group definitions} \label{tbl:uploads_alt_control}
 {\small
\input{Overleaf/tables/new/uploads_alt_control}\\
}
%\noalign{\vskip .5em}
\begin{minipage}{\linewidth}
{\footnotesize
\def\sym#1{\ifmmode^{#1}\else\(^{#1}\)\fi}
\textbf{Note:} 
The dependent variable in columns (1) and (4) is the total number of uploads per month, in columns (2) and (5) it is an indicator of whether a user account has uploaded at least one image in a given month, and it is the log number of uploads in columns (3) and (6). The control group in columns (1)--(3) includes users that have uploaded at least one non-curated and non-nature themed image. The control group in columns (4)--(6) includes all users that do not appear in the LITE data. \textit{Treated} indicates whether a user was part of the LITE dataset. Fixed effects for upload month of the image and user account. Standard errors clustered at the user level in parentheses. \sym{*} \(p<0.10\), \sym{**} \(p<0.05\) \sym{***} \(p<0.01\)
}
\end{minipage}
\end{table}

\begin{table}[!h]
 \caption{Results: Changes in uploads, alternative control group definitions (weighted)} \label{tbl:uploads_alt_control_weighted}
 {\small
\input{Overleaf/tables/new/uploads_alt_control_weighted}\\
}
%\noalign{\vskip .5em}
\begin{minipage}{\linewidth}
{\footnotesize
\def\sym#1{\ifmmode^{#1}\else\(^{#1}\)\fi}
\textbf{Note:} 
The dependent variable in columns (1) and (4) is the total number of uploads per month, in columns (2) and (5) it is an indicator of whether a user account has uploaded at least one image in a given month, and it is the log number of uploads in columns (3) and (6). The control group in columns (1)--(3) includes users that have uploaded at least one non-curated and non-nature themed image. The control group in columns (4)--(6) includes all users that do not appear in the LITE data. \textit{Treated} indicates whether a user was part of the LITE dataset. Fixed effects for upload month of the image and user account. Standard errors clustered at the user level in parentheses. \sym{*} \(p<0.10\), \sym{**} \(p<0.05\) \sym{***} \(p<0.01\)
}
\end{minipage}
\end{table}

\begin{table}[!h]
 \caption{Results: Changes in uploads, Poisson models} \label{tbl:uploads_user_percentiles_ppml}
\input{Overleaf/tables/new/uploads_user_percentiles_ppml}\\
%\noalign{\vskip .5em}
\begin{minipage}{\linewidth}
{\footnotesize
\def\sym#1{\ifmmode^{#1}\else\(^{#1}\)\fi}
\textbf{Note:} We estimate Poisson pseudo-maximum likelihood regressions models. The dependent variable is the number of uploads in a given month. \textit{Treated} indicates whether a user was part of the LITE dataset. \textit{Above99thPct}, \textit{Above95thPct}, and \textit{Above90thPct} indicate whether a user's uploads were above the 99/95/90th percentile. The samples in columns (3) to (5) do not include users above the 99/95/90th percentile of the pre-treatment uploads. The number of observations is smaller than in Tables \ref{tbl:uploads} and \ref{tbl:uploads_user_percentiles} because singleton observations are dropped due to fixed effects. Standard errors clustered at the user level in parentheses. \sym{*} \(p<0.10\), \sym{**} \(p<0.05\) \sym{***} \(p<0.01\)  \sym{*} \(p<0.10\), \sym{**} \(p<0.05\) \sym{***} \(p<0.01\)
}
\end{minipage}
\end{table}

\begin{table}[!t]
 \caption{Results: Changes in uploads, by user activity} \label{tbl:uploads_user_percentiles}
\input{Overleaf/tables/new/uploads_user_percentiles}\\
%\noalign{\vskip .5em}
\begin{minipage}{\linewidth}
{\footnotesize
\def\sym#1{\ifmmode^{#1}\else\(^{#1}\)\fi}
\textbf{Note:} 
The dependent variable is the number of uploads in a given month. \textit{Treated} indicates whether a user was part of the LITE dataset. \textit{Above99thPct}, \textit{Above95thPct} and \textit{Above90thPct} indicate whether the users' uploads were above the 99/95/90th percentile. The samples in columns (2) to (4) do not include users above the 99/95/90th percentile of uploads. \sym{*} \(p<0.10\), \sym{**} \(p<0.05\) \sym{***} \(p<0.01\)
}
\end{minipage}
\end{table}

\end{document}

%% file: Overleaf/tables/order.tex
{
\def\sym#1{\ifmmode^{#1}\else\(^{#1}\)\fi}
\begin{tabular}{l*{2}{rrrrrr}}
\toprule
                &\multicolumn{1}{c}{(1)}         &\multicolumn{1}{c}{(2)}         \\
\midrule
AutoKeywordScore&  -3.3952\sym{***}&  -3.5100\sym{***}\\
                & (1.1170)         & (1.1116)         \\
\addlinespace
ImagePopularity &  -0.0000         &  -0.0000         \\
                & (0.0000)         & (0.0000)         \\
\addlinespace
UserKeyword     & 136.1898         & 139.4685         \\
                &(119.9305)         &(121.0681)         \\
\addlinespace
ImageAge        & -25.7948         &  12.3496         \\
                &(37.7172)         &(48.5456)         \\
\addlinespace
UserPopularity  &                  &   0.1280         \\
                &                  & (0.1381)         \\
\addlinespace
AccountAge      &                  & -55.0069         \\
                &                  &(42.8323)         \\
\addlinespace
TotalUploads    &                  &  -0.0385         \\
                &                  & (0.0588)         \\
\midrule
adj. R2         &   0.0003         &   0.0003         \\
Observations    &   25,000         &   24,964         \\
\bottomrule
\end{tabular}
}

%% file: Overleaf/tables/new/survival_image_user.tex
{
\def\sym#1{\ifmmode^{#1}\else\(^{#1}\)\fi}
\begin{tabular}{l*{5}{rrrrrr}}
\toprule
                &\multicolumn{3}{c}{Images}                              &\multicolumn{2}{c}{Users}            \\\cmidrule(lr){2-4}\cmidrule(lr){5-6}
                &\multicolumn{1}{c}{(1)}         &\multicolumn{1}{c}{(2)}         &\multicolumn{1}{c}{(3)}         &\multicolumn{1}{c}{(4)}         &\multicolumn{1}{c}{(5)}         \\
\midrule
Treated         &   0.0220\sym{**} &   0.0219\sym{**} &   0.0109\sym{***}&  -0.0075\sym{*}  &  -0.0084\sym{**} \\
                &(0.00875)         &(0.00879)         &(0.00155)         &(0.00395)         &(0.00406)         \\
\midrule
Mean Control    &   0.9075         &   0.9075         &   0.9075         &   0.9624         &   0.9625         \\
Image-Age FE    &      Yes         &      Yes         &      Yes         &      Yes         &      Yes         \\
User FE         &       No         &       No         &      Yes         &       No         &       No         \\
Observations    &  705,268         &  704,848         &  704,848         &   12,033         &   11,607         \\
\bottomrule
\end{tabular}
}

%% file: Overleaf/tables/new/uploads_fe.tex
{
\def\sym#1{\ifmmode^{#1}\else\(^{#1}\)\fi}
\begin{tabular}{l*{6}{rrrrrr}}
\toprule
                &\multicolumn{2}{c}{Uploads}          &\multicolumn{2}{c}{I(Uploads$>$0)}   &\multicolumn{2}{c}{Log(1+Uploads)}   \\\cmidrule(lr){2-3}\cmidrule(lr){4-5}\cmidrule(lr){6-7}
                &\multicolumn{1}{c}{(1)}         &\multicolumn{1}{c}{(2)}         &\multicolumn{1}{c}{(3)}         &\multicolumn{1}{c}{(4)}         &\multicolumn{1}{c}{(5)}         &\multicolumn{1}{c}{(6)}         \\
\midrule
Post $\times$ Treated&  -1.0623\sym{***}&  -1.1099\sym{***}&  -0.0694\sym{***}&  -0.0691\sym{***}&  -0.1605\sym{***}&  -0.1630\sym{***}\\
                &(0.09318)         &(0.09693)         &(0.00354)         &(0.00352)         &(0.00717)         &(0.00728)         \\
\midrule
Mean Treated Before&   2.9038         &   2.9038         &   0.2635         &   0.2635         &   0.4892         &   0.4892         \\
Month FE        &      Yes         &      Yes         &      Yes         &      Yes         &      Yes         &      Yes         \\
User FE         &       No         &      Yes         &       No         &      Yes         &       No         &      Yes         \\
Observations    &  612,662         &  612,662         &  612,662         &  612,662         &  612,662         &  612,662         \\
\bottomrule
\end{tabular}
}

%% file: Overleaf/tables/new/uploads_user_types.tex
{
\def\sym#1{\ifmmode^{#1}\else\(^{#1}\)\fi}
\begin{tabular}{l*{2}{rrrrrr}}
\toprule
                &\multicolumn{1}{c}{(1)}         &\multicolumn{1}{c}{(2)}         \\
\midrule
Post $\times$ Treated&  -0.6740\sym{***}&  -0.7206\sym{***}\\
                &(0.13433)         &(0.09401)         \\
\addlinespace
Post $\times$ Pro-Gear&  -0.0686         &                  \\
                &(0.10904)         &                  \\
\addlinespace
Post $\times$ Treated $\times$ Pro-Gear&  -0.5038\sym{***}&                  \\
                &(0.17407)         &                  \\
\addlinespace
Post $\times$ ForHire&                  &  -0.2362         \\
                &                  &(0.16669)         \\
\addlinespace
Post $\times$ Treated $\times$ ForHire&                  &  -0.8917\sym{***}\\
                &                  &(0.25200)         \\
\midrule
Mean DV         &   1.4595         &   1.4474         \\
Month FE        &      Yes         &      Yes         \\
User FE         &      Yes         &      Yes         \\
Observations    &  611,891         &  577,739         \\
\bottomrule
\end{tabular}
}

%% file: Overleaf/tables/new/uploads_mechanisms.tex
{
\def\sym#1{\ifmmode^{#1}\else\(^{#1}\)\fi}
\begin{tabular}{l*{2}{rrrrrr}}
\toprule
                &\multicolumn{2}{c}{Uploads}          \\\cmidrule(lr){2-3}
                &\multicolumn{1}{c}{(1)}         &\multicolumn{1}{c}{(2)}         \\
\midrule
Post $\times$ TreatedSingle&  -0.3059\sym{***}&                  \\
                &(0.09750)         &                  \\
\addlinespace
Post $\times$ TreatedMultiple&  -2.2830\sym{***}&                  \\
                &(0.17248)         &                  \\
\addlinespace
Post $\times$ Treated&                  &  -1.0056\sym{***}\\
                &                  &(0.10508)         \\
\addlinespace
PostAug22 $\times$ Treated&                  &  -0.4054\sym{***}\\
                &                  &(0.10056)         \\
\midrule
Mean Treated Before&                  &   2.9038         \\
Mean TreatedSingle Before&   1.4011         &                  \\
Mean TreatedMultiple Before&   5.0982         &                  \\
Month FE        &      Yes         &      Yes         \\
User FE         &      Yes         &      Yes         \\
Observations    &  612,662         &  612,662         \\
\bottomrule
\end{tabular}
}

%% file: Overleaf/tables/new/uploads_image_types.tex
{
\def\sym#1{\ifmmode^{#1}\else\(^{#1}\)\fi}
\begin{tabular}{l*{3}{rrrrrr}}
\toprule
                &\multicolumn{1}{c}{(1)}&\multicolumn{1}{c}{(2)}&\multicolumn{1}{c}{(3)}\\
                &\multicolumn{1}{c}{Nature}&\multicolumn{1}{c}{Curated}&\multicolumn{1}{c}{Nature\&Curated}\\
\midrule
Post $\times$ Treated&  -0.2265\sym{***}&  -0.3472\sym{***}&  -0.1392\sym{***}\\
                &(0.02956)         &(0.01642)         &(0.00614)         \\
\midrule
Mean Treated Before&   0.7218         &   0.5739         &   0.2022         \\
Month FE        &      Yes         &      Yes         &      Yes         \\
User FE         &      Yes         &      Yes         &      Yes         \\
Observations    &  612,662         &  612,662         &  612,662         \\
\bottomrule
\end{tabular}
}

%% file: Overleaf/tables/new/instagram.tex
{
\def\sym#1{\ifmmode^{#1}\else\(^{#1}\)\fi}
\begin{tabular}{l*{4}{rrrrrr}}
\toprule
                &\multicolumn{1}{c}{Unsplash}&\multicolumn{1}{c}{Instagram}&\multicolumn{2}{c}{Both}             \\\cmidrule(lr){2-2}\cmidrule(lr){3-3}\cmidrule(lr){4-5}
                &\multicolumn{1}{c}{(1)}         &\multicolumn{1}{c}{(2)}         &\multicolumn{1}{c}{(3)}         &\multicolumn{1}{c}{(4)}         \\
\midrule
                                                                                            \\
Treated $\times$ Post&  -0.5118\sym{***}&  -0.0013         &  -0.4344\sym{***}&  -0.5111\sym{***}\\
                &(0.08951)         &(0.02326)         &(0.06900)         &(0.06537)         \\
Instagram       &                  &                  &   0.0704         &             \\
                &                  &                  &(0.06441)         &                \\
Treated $\times$ Instagram&                  &                  &  -0.7933\sym{***}&            \\
                &                  &                  &(0.07820)         &                \\
Post $\times$ Instagram&                  &                  &   0.0003         &  -0.0093         \\
                &                  &                  &(0.07802)         &(0.07604)         \\
Treated $\times$ Post $\times$ Instagram&                  &                  &   0.3556\sym{***}&   0.5090\sym{***}\\
                &                  &                  &(0.09457)         &(0.09240)         \\
Constant        &   0.9318\sym{***}&   0.3874\sym{***}&   0.8952\sym{***}&   0.6627\sym{***}\\
                &(0.04588)         &(0.01192)         &(0.03553)         &(0.03525)         \\
\midrule
Total effect Instagram                                                                      \\
lincom          &                  &                  &  -0.0787         &  -0.0020         \\
                &                  &                  &(0.06900)         &(0.06537)         \\
\midrule
Mean DV         &   1.4300         &   0.7071         &   1.0686         &   1.0686         \\
Month FE        &      Yes         &      Yes         &      Yes         &      Yes         \\
User FE         &      Yes         &      Yes         &      Yes         &       No         \\
User-Platform FE&       No         &       No         &       No         &      Yes         \\
Observations    &   56,855         &   56,855         &  113,728         &  113,728         \\
\bottomrule
\end{tabular}
}

%% file: Overleaf/tables/new/uploads_image_sim.tex
{
\def\sym#1{\ifmmode^{#1}\else\(^{#1}\)\fi}
\begin{tabular}{l*{6}{rrrrrr}}
\toprule
                &\multicolumn{2}{c}{Avg. Similarity}  &\multicolumn{2}{c}{LogVerySimilar (0.8)}&\multicolumn{2}{c}{LogVerySimilar (0.9)}\\\cmidrule(lr){2-3}\cmidrule(lr){4-5}\cmidrule(lr){6-7}
                &\multicolumn{1}{c}{(1)}         &\multicolumn{1}{c}{(2)}         &\multicolumn{1}{c}{(3)}         &\multicolumn{1}{c}{(4)}         &\multicolumn{1}{c}{(5)}         &\multicolumn{1}{c}{(6)}         \\
\midrule
Post $\times$ Treated&   0.0189\sym{***}&   0.0070\sym{**} &   0.2520\sym{***}&   0.0749         &   0.1336\sym{***}&   0.0380         \\
                &(0.00097)         &(0.00345)         &(0.01817)         &(0.05707)         &(0.02300)         &(0.06871)         \\
\midrule
Mean Treated Before&   0.3910         &   0.3910         &   8.1285         &   8.1285         &   4.5527         &   4.5527         \\
Image-Age FE    &      Yes         &      Yes         &      Yes         &      Yes         &      Yes         &      Yes         \\
User FE         &       No         &      Yes         &       No         &      Yes         &       No         &      Yes         \\
Observations    &  340,991         &  340,346         &  340,991         &  340,346         &  340,991         &  340,346         \\
\bottomrule
\end{tabular}
}

%% file: Overleaf/tables/new/survival_image_usertypes.tex
{
\def\sym#1{\ifmmode^{#1}\else\(^{#1}\)\fi}
\begin{tabular}{l*{3}{rrrrrr}}
\toprule
                &\multicolumn{3}{c}{Images}                              \\\cmidrule(lr){2-4}
                &\multicolumn{1}{c}{(1)}         &\multicolumn{1}{c}{(2)}         &\multicolumn{1}{c}{(3)}         \\
\midrule
Treated         &   0.0109\sym{***}&   0.0226\sym{***}&   0.0203\sym{***}\\
                &(0.00155)         &(0.00364)         &(0.00401)         \\
\addlinespace
Treated $\times$ Pro-Gear&                  &  -0.0051         &  -0.0067\sym{*}  \\
                &                  &(0.00329)         &(0.00345)         \\
\addlinespace
Treated $\times$ AccountAge&                  &  -0.0003\sym{***}&  -0.0002\sym{***}\\
                &                  &(0.00008)         &(0.00008)         \\
\addlinespace
Treated $\times$ Multiple&                  &  -0.0066\sym{***}&  -0.0071\sym{***}\\
                &                  &(0.00243)         &(0.00241)         \\
\addlinespace
Treated $\times$ ForHire&                  &                  &   0.0091\sym{***}\\
                &                  &                  &(0.00317)         \\
\midrule
Mean Control    &   0.9075         &   0.9075         &   0.9075         \\
Image-Age FE    &      Yes         &      Yes         &      Yes         \\
User FE         &      Yes         &      Yes         &      Yes         \\
Observations    &  704,848         &  704,848         &  648,653         \\
\bottomrule
\end{tabular}
}

%% file: Overleaf/tables/new/uploads_weighted.tex
{
\def\sym#1{\ifmmode^{#1}\else\(^{#1}\)\fi}
\begin{tabular}{l*{3}{rrrrrr}}
\toprule
                &\multicolumn{1}{c}{Uploads}&\multicolumn{1}{c}{I(Uploads$>$0)}&\multicolumn{1}{c}{Log(1+Uploads)}\\\cmidrule(lr){2-2}\cmidrule(lr){3-3}\cmidrule(lr){4-4}
                &\multicolumn{1}{c}{(1)}         &\multicolumn{1}{c}{(2)}         &\multicolumn{1}{c}{(3)}         \\
\midrule
Post $\times$ Treated&  -1.0573\sym{***}&  -0.0685\sym{***}&  -0.1599\sym{***}\\
                &(0.08746)         &(0.00351)         &(0.00715)         \\
\midrule
Mean Treated Before&   2.9038         &   0.2635         &   0.4892         \\
Month FE        &      Yes         &      Yes         &      Yes         \\
User FE         &      Yes         &      Yes         &      Yes         \\
Observations    &  612,662         &  612,662         &  612,662         \\
\bottomrule
\end{tabular}
}

%% file: Overleaf/tables/new/uploads_alt_control.tex
{
\def\sym#1{\ifmmode^{#1}\else\(^{#1}\)\fi}
\begin{tabular}{l*{6}{rrrrrr}}
\toprule
                &\multicolumn{3}{c}{Not Nature\&Curated}                 &\multicolumn{3}{c}{Not in LITE}                         \\\cmidrule(lr){2-4}\cmidrule(lr){5-7}
                &\multicolumn{1}{c}{(1)}&\multicolumn{1}{c}{(2)}&\multicolumn{1}{c}{(3)}&\multicolumn{1}{c}{(4)}&\multicolumn{1}{c}{(5)}&\multicolumn{1}{c}{(6)}\\
                &\multicolumn{1}{c}{Uploads}&\multicolumn{1}{c}{I(Upl.$>$0)}&\multicolumn{1}{c}{Log(1+Upl.)}&\multicolumn{1}{c}{Uploads}&\multicolumn{1}{c}{I(Upl.$>$0)}&\multicolumn{1}{c}{Log(1+Upl.)}\\
\midrule
Post $\times$ Treated&  -1.3737\sym{***}&  -0.0857\sym{***}&  -0.2072\sym{***}&  -1.4495\sym{***}&  -0.1041\sym{***}&  -0.2323\sym{***}\\
                &(0.08534)         &(0.00238)         &(0.00539)         &(0.08527)         &(0.00236)         &(0.00537)         \\
\midrule
Mean Treated Before&   2.9038         &   0.2635         &   0.4892         &   2.9038         &   0.2635         &   0.4892         \\
Month FE        &      Yes         &      Yes         &      Yes         &      Yes         &      Yes         &      Yes         \\
User FE         &      Yes         &      Yes         &      Yes         &      Yes         &      Yes         &      Yes         \\
Observations    &7,666,123         &7,666,123         &7,666,123         &9,704,756         &9,704,756         &9,704,756         \\
\bottomrule
\end{tabular}
}

%% file: Overleaf/tables/new/uploads_alt_control_weighted.tex
{
\def\sym#1{\ifmmode^{#1}\else\(^{#1}\)\fi}
\begin{tabular}{l*{6}{rrrrrr}}
\toprule
                &\multicolumn{3}{c}{Not Nature\&Curated}                 &\multicolumn{3}{c}{Not in LITE}                         \\\cmidrule(lr){2-4}\cmidrule(lr){5-7}
                &\multicolumn{1}{c}{(1)}&\multicolumn{1}{c}{(2)}&\multicolumn{1}{c}{(3)}&\multicolumn{1}{c}{(4)}&\multicolumn{1}{c}{(5)}&\multicolumn{1}{c}{(6)}\\
                &\multicolumn{1}{c}{Uploads}&\multicolumn{1}{c}{I(Upl.$>$0)}&\multicolumn{1}{c}{Log(1+Upl.)}&\multicolumn{1}{c}{Uploads}&\multicolumn{1}{c}{I(Upl.$>$0)}&\multicolumn{1}{c}{Log(1+Upl.)}\\
\midrule
Post $\times$ Treated&  -1.3176\sym{***}&  -0.0849\sym{***}&  -0.2037\sym{***}&  -1.3930\sym{***}&  -0.1033\sym{***}&  -0.2287\sym{***}\\
                &(0.07575)         &(0.00236)         &(0.00525)         &(0.07567)         &(0.00235)         &(0.00524)         \\
\midrule
Mean Treated Before&   0.4892         &   0.2635         &   0.4892         &   0.4892         &   0.2635         &   0.4892         \\
Month FE        &      Yes         &      Yes         &      Yes         &      Yes         &      Yes         &      Yes         \\
User FE         &      Yes         &      Yes         &      Yes         &      Yes         &      Yes         &      Yes         \\
Observations    &7,666,123         &7,666,123         &7,666,123         &9,704,756         &9,704,756         &9,704,756         \\
\bottomrule
\end{tabular}
}

%% file: Overleaf/tables/new/uploads_user_percentiles_ppml.tex
{
\def\sym#1{\ifmmode^{#1}\else\(^{#1}\)\fi}
\begin{tabular}{l*{5}{rrrrrr}}
\toprule
                &\multicolumn{2}{c}{All}              &\multicolumn{1}{c}{w/o 99th}&\multicolumn{1}{c}{w/o 95th}&\multicolumn{1}{c}{w/o 90th}\\\cmidrule(lr){2-3}\cmidrule(lr){4-4}\cmidrule(lr){5-5}\cmidrule(lr){6-6}
                &\multicolumn{1}{c}{(1)}         &\multicolumn{1}{c}{(2)}         &\multicolumn{1}{c}{(3)}         &\multicolumn{1}{c}{(4)}         &\multicolumn{1}{c}{(5)}         \\
\midrule
Post $\times$ Treated&   0.0607         &  -0.1996\sym{***}&  -0.0541         &  -0.1703\sym{***}&  -0.2008\sym{***}\\
                &(0.12357)         &(0.05497)         &(0.06611)         &(0.05785)         &(0.05866)         \\
\addlinespace
Post $\times$ Treated $\times$ Above99thPct&                  &  -0.3154         &                  &                  &                  \\
                &                  &(0.38920)         &                  &                  &                  \\
\addlinespace
Post $\times$ Treated $\times$ Above95thPct&                  &   0.1842         &                  &                  &                  \\
                &                  &(0.24548)         &                  &                  &                  \\
\addlinespace
Post $\times$ Treated $\times$ Above90thPct&                  &  -0.2198         &                  &                  &                  \\
                &                  &(0.16234)         &                  &                  &                  \\
\midrule
Mean Treated Before&   2.9038         &   2.9038         &   2.2160         &   1.5488         &   1.1435         \\
Month FE        &      Yes         &      Yes         &      Yes         &      Yes         &      Yes         \\
User FE         &      Yes         &      Yes         &      Yes         &      Yes         &      Yes         \\
Observations    &  468,555         &  468,555         &  462,383         &  437,881         &  406,969         \\
\bottomrule
\end{tabular}
}

%% file: Overleaf/tables/new/uploads_user_percentiles.tex
{
\def\sym#1{\ifmmode^{#1}\else\(^{#1}\)\fi}
\begin{tabular}{l*{4}{rrrrrr}}
\toprule
                &\multicolumn{1}{c}{(1)}&\multicolumn{1}{c}{(2)}&\multicolumn{1}{c}{(3)}&\multicolumn{1}{c}{(4)}\\
                &\multicolumn{1}{c}{All}&\multicolumn{1}{c}{w/o 99th}&\multicolumn{1}{c}{w/o 95th}&\multicolumn{1}{c}{w/o 90th}\\
\midrule
Post $\times$ Treated&  -0.4529\sym{***}&  -0.9872\sym{***}&  -0.6847\sym{***}&  -0.4571\sym{***}\\
                &(0.02672)         &(0.05445)         &(0.03513)         &(0.02699)         \\
\addlinespace
Post $\times$ Treated $\times$ Above99thPct& -17.4338         &                  &                  &                  \\
                &(11.68857)         &                  &                  &                  \\
\addlinespace
Post $\times$ Treated $\times$ Above95thPct&  -0.9941         &                  &                  &                  \\
                &(2.04651)         &                  &                  &                  \\
\addlinespace
Post $\times$ Treated $\times$ Above90thPct&  -1.3862\sym{*}  &                  &                  &                  \\
                &(0.74521)         &                  &                  &                  \\
\midrule
Mean Treated Before&   2.9038         &   2.2160         &   1.5488         &   1.1435         \\
Month FE        &      Yes         &      Yes         &      Yes         &      Yes         \\
User FE         &      Yes         &      Yes         &      Yes         &      Yes         \\
Observations    &  612,662         &  606,490         &  581,988         &  551,076         \\
\bottomrule
\end{tabular}
}